\documentclass[a4paper,12pt]{article}
\usepackage{siunitx}
\usepackage{graphicx}
\usepackage{afterpage}
\usepackage{amsmath}
\usepackage{amsfonts}
\usepackage{amssymb}
\usepackage{parskip}
\usepackage{bm}
\usepackage{url}
\usepackage{booktabs}
\usepackage{multirow}
\usepackage{chngcntr}
\usepackage{tikz}
\usetikzlibrary{bayesnet}

\usepackage[margin=1.0in]{geometry}
\usepackage{authblk}
\usepackage[nopar]{lipsum}
\usepackage[hang,flushmargin]{footmisc}
\newcommand\blfootnote[1]{%
	\begingroup
	\renewcommand\thefootnote{}\footnote{#1}%
	\addtocounter{footnote}{-1}%
	\endgroup
}
\usepackage{caption}
\usepackage{natbib}
\title{Bayesian hierarchical factor regression models to infer cause of death from verbal autopsy data}

\author[1]{Kelly R. Moran}
\author[1]{Elizabeth L. Turner}
\author[1]{David Dunson}
\author[1]{Amy H. Herring}

\affil[1]{Duke University, Durham, USA}

\date{}

\begin{document}
	
\maketitle

\renewcommand{\abstractname}{Summary}
\begin{abstract}
In low-resource settings where vital registration of death is not routine it is often of critical interest to determine and study the cause of death (COD) for individuals and the cause-specific mortality fraction (CSMF) for populations. Post-mortem autopsies, considered the gold standard for COD assignment, are often difficult or impossible to implement due to deaths occurring outside the hospital, expense, and/or cultural norms. For this reason, Verbal Autopsies (VAs) are commonly conducted, consisting of a questionnaire administered to next of kin recording demographic information, known medical conditions, symptoms, and other factors for the decedent. This article proposes a novel class of hierarchical factor regression models that avoid restrictive assumptions of standard methods, allow both the mean and covariance to vary with COD category, and can include covariate information on the decedent, region, or events surrounding death. Taking a Bayesian approach to inference, this work develops an MCMC algorithm and validates the FActor Regression for Verbal Autopsy (FARVA) model in simulation experiments. An application of FARVA to real VA data shows improved goodness-of-fit and better predictive performance in inferring COD and CSMF over competing methods. Code and a user manual are made available at \url{https://github.com/kelrenmor/farva}.

\vspace{1\baselineskip}\vspace{-\parskip}
	
\end{abstract}


\pagebreak 

\section{Introduction}

When it comes to the global burden of disease, the weight falls most heavily on the shoulders of low-income countries. As measured by Disability Adjusted Life Years (DALYs) lost, estimated global rates range from 40,000 to 70,000 DALYs per 100,000 individuals across low-income countries. On the other hand, rates in most developed countries tend to fall between 10,000 and 30,000 DALYs per 100,000 individuals \citep{GBDCN2017, owidburdenofdisease}. Furthermore, many deaths in low-income countries occur without registration, recording, or notice by the health system \citep{nichols20182016}. Hospitals, community health workers, and public health planners are hindered in their ability to treat new patients, allocate resources, and plan for the future when the landscape of cause of death (COD) is poorly mapped.

Medical certification of cause of death without an autopsy is difficult in most cases. Unfortunately, performing an autopsy is often infeasible or impossible. Decedents often pass away in the home. If deaths occur in the hospital, the next of kin may not agree to a full or even minimally invasive autopsy. If the next of kin does agree, the hospital may lack the resources to perform the autopsy, or the findings may be inconclusive. In cases where no medical certification occurs, verbal autopsy (VA) offers a practical alternative approach for assessing cause of death. VA involves exploring the signs and symptoms (hereafter referred to only by symptoms) a decedent experienced before death by structured interview with a relative or caregiver of the deceased. Questions can include details such as ``For how long was [decedent name] ill before they died?'' and ``Did [decedent name] have a fever in the three weeks leading up to death?'' The 2016 World Health Organization (WHO) VA instrument \citep{va2016}, available online at \url{https://www.who.int/healthinfo/statistics/verbalautopsystandards/en/}, offers a standardized form by which VA data may be recorded.

One option for analyzing these VA records is to have physicians go through and decide on a COD for each decedent. However, this process is time consuming and takes resources away from existing patients in already resource-limited settings. In addition, such expert labeling may differ when multiple physicians are presented with the same data. An alternative approach is computer-coded VA, in which the COD is assigned via an algorithm or probabilistic model. Existing computer-coded VA algorithms include those for which the relationship between symptoms and cause of death is encoded by experts (InterVA \citep{byass2012strengthening} and InSilicoVA \citep{mccormick2016probabilistic}) and those for which it is learned by relying on a labeled subset of the data having known COD (the King and Lu method \citep{king2008verbal}, the Tariff method \citep{james2011performance}, the Simplified Symptom Pattern method \citep{murray2011simplified}, the naive Bayes classifier \citep{miasnikof2015naive}, the Bayesian factor model \citep{kunihama2018bayesian}, and latent Gaussian graphical model \citep{li2018using}).

The majority of the work on VA algorithms relies on a conditional independence assumption \citep{mccormick2016probabilistic, byass2012strengthening, james2011performance, miasnikof2015naive}. That is, assume that once the cause of death is known, the knowledge that an individual had symptom $A$ should not impact belief that an individual had symptom $B$ for ANY combination of symptoms $A$ and $B$. Recent work \citep{li2018using, kunihama2018bayesian} has shown that this assumption is not valid in general. Logically, this finding is unsurprising; for example, the knowledge that someone had difficulty breathing would increase belief that they had a cough even given that their COD is already known.

\citet{li2018using} and \citet{kunihama2018bayesian} have relaxed the conditional independence assumption by probabilistically addressing conditional associations of symptoms given causes. The Bayesian latent Gaussian graphical model of \citet{li2018using} assumes the conditional dependence structure is common across causes, i.e., the symptom-level correlation does not vary with cause. On the other hand, the latent factor model of \citet{kunihama2018bayesian} models the symptom-level association independently for each cause. Neither allows for covariates (e.g. age, season, malaria endemicity of region) to be included in the model unless they are treated as ``symptoms'' themselves. The capability to allow covariates to affect cause assignment is potentially quite useful. For example, the information that an individual had balance and memory issues may be highly informative for certain CODs given an individual is young, but be uninformative among elder decedents for whom such symptoms are quite common.

\begin{figure}[!htb]
	\centering
	\includegraphics[width=1.0\textwidth,keepaspectratio]{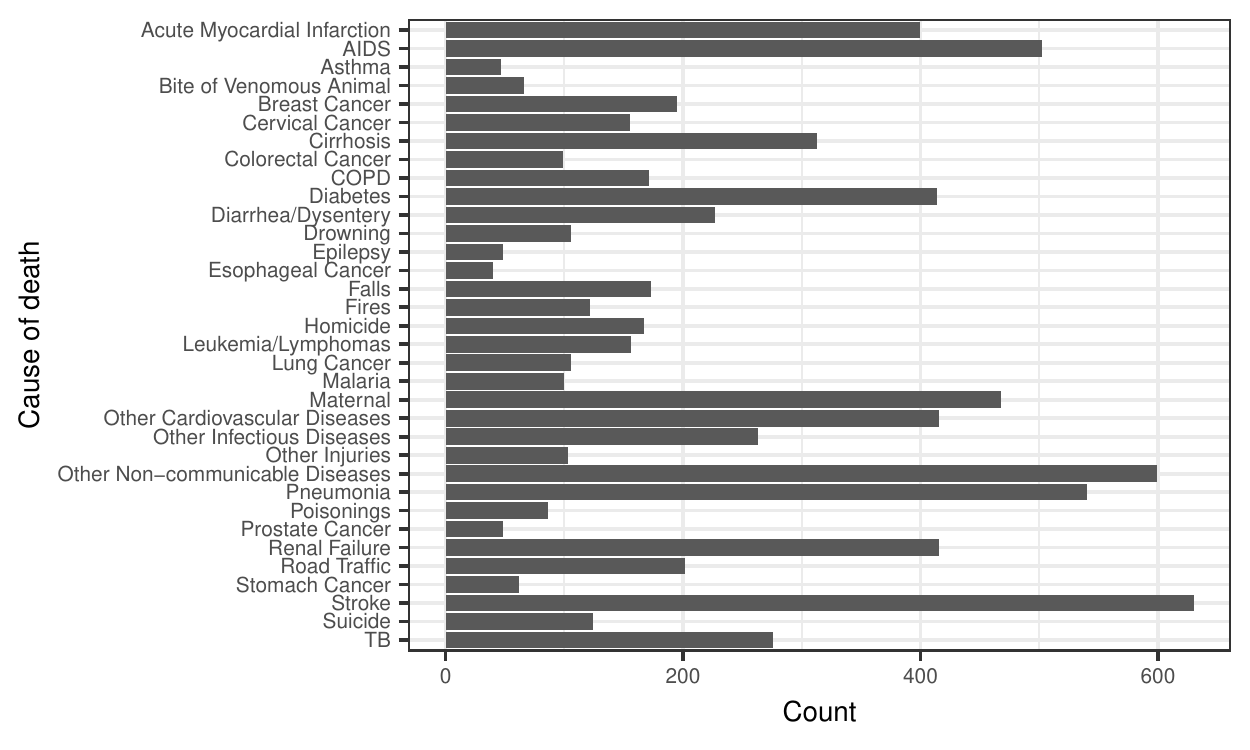}
	\caption{Number of deaths due to each of the 34 broad cause groupings in the adult PHMRC data set ($n=7840$). See the online supplementary materials for an analogous plot broken down by age.}
	\label{fig:phmrc_causes}
\end{figure}

The Population Health Metrics Research Consortium (PHMRC) created an open source ``Gold Standard'' VA database for training and testing VA models \citep{murray2011population}. This database uses a standardized VA questionnaire developed by the PHMRC based on WHO standards. These data include 7,840 ``adults'' (defined by both the WHO and PHMRC questionnaires to mean decedents aged 12 and older), for whom potential causes for analysis number 34 for adults (see Figure \ref{fig:phmrc_causes}).  Locations include six sites across four countries, and data were collected from 2007-2010. The ensuing examination of the PHMRC data illustrates the importance of allowing both the mean and the covariance to vary with both COD and covariate information, highlighting the importance of the methodological development in this article.

\begin{figure}[!htb]
	\centering
	\includegraphics[width=1.0\textwidth,keepaspectratio]{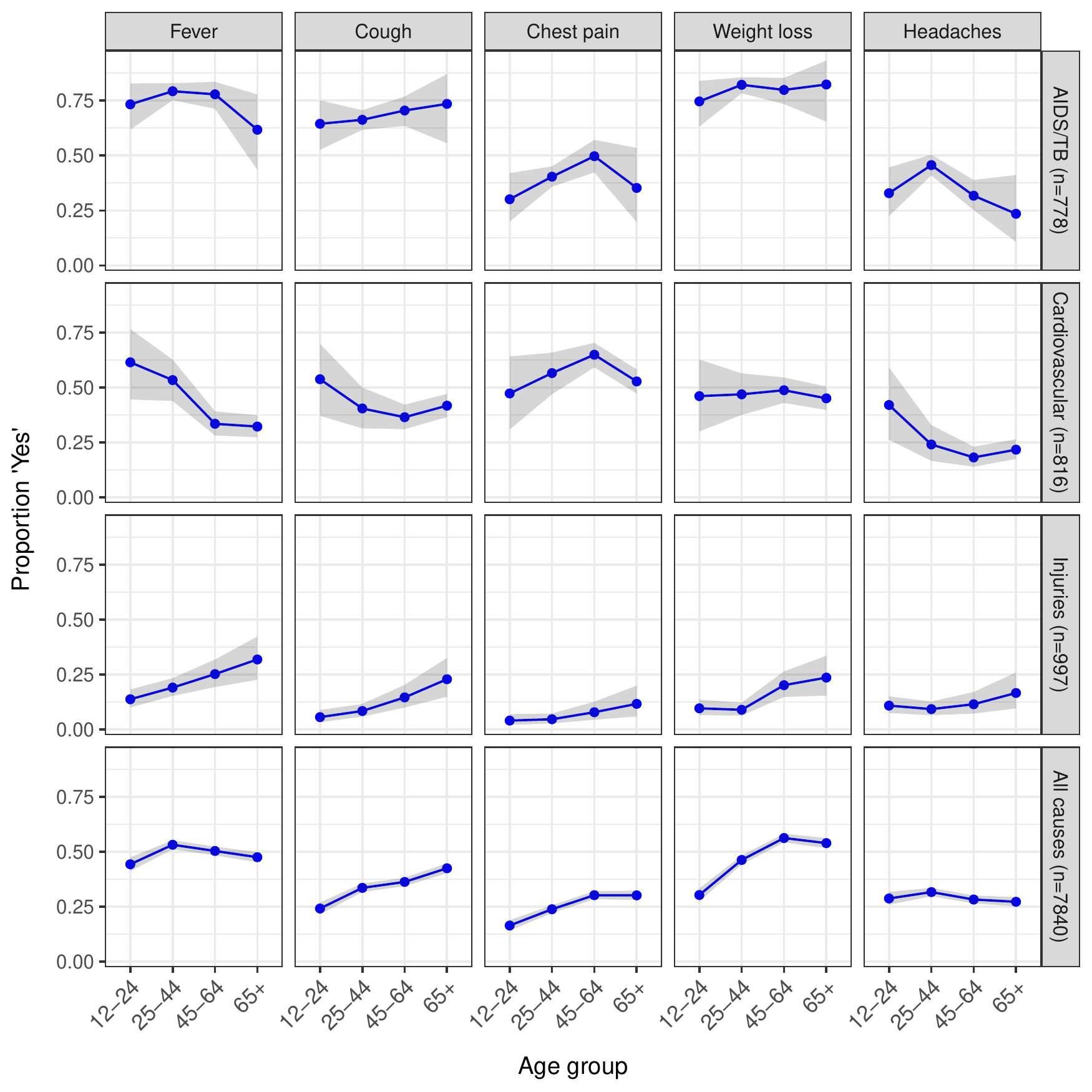}
	\caption{The average proportion of symptoms present varies across both age and cause of death (COD) in the adult PHMRC data. Each subplot corresponds to the proportion of ``Yes'' responses given by interviewees when asked about each symptom with proportion point estimates (points) and 95\% confidence intervals (bands) shown. Responses of don't know or refuse to answer are omitted, and individuals having unknown age (123) are omitted from age-specific subplots. A more detailed figure showing the number of observations contributing to each data point is provided in the online supplementary materials.} 
	\label{fig:symptom_mean}
\end{figure}

\begin{figure}[hbpt]
	\centering
	\includegraphics[width=1.0\textwidth,keepaspectratio]{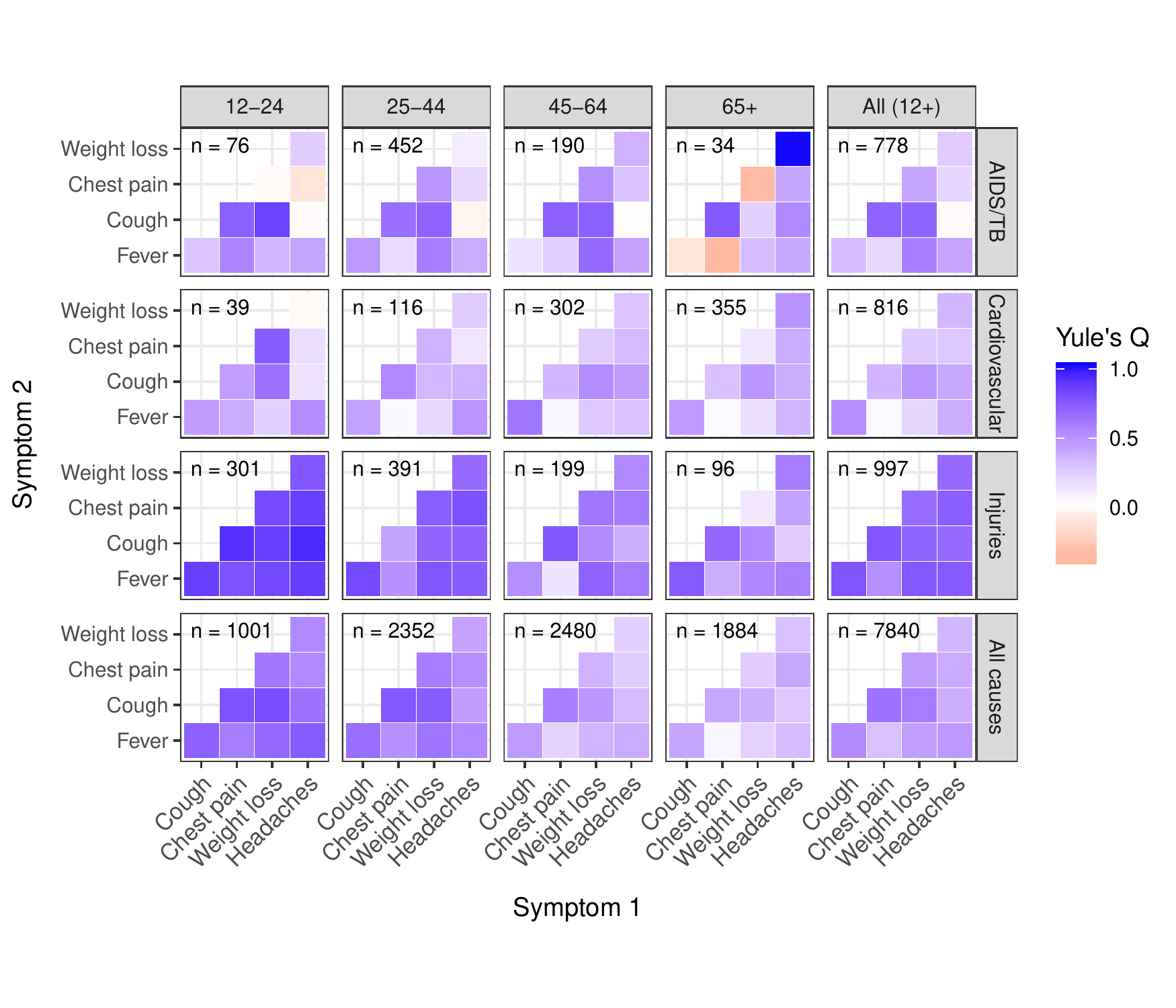}
	\caption{The association between symptoms varies across both age and cause of death (COD). Each subplot corresponds to the association between pairs of select symptoms in the adult PHMRC data as measured by Yule's Q (possible range $-1$ to $1$). Rows correspond to COD groups and columns correspond to age groups. ``Don't know'' or ``refused to answer'' responses were omitted for the Yule's Q calculations, and individuals having unknown age (123) are omitted from age-specific subplots. A more detailed figure showing the number of observations contributing to the calculations in each subplot symptom-by-symptom square is provided in the online supplementary materials.}
	\label{fig:symptom_cov} 
\end{figure}

Figures \ref{fig:symptom_mean} and \ref{fig:symptom_cov} offer illustrative data-driven examples from the adult PHMRC data set using the five symptoms of fever, cough, chest pain, weight loss, and headaches and the three broad COD clusters AIDS/TB, cardiovascular conditions (including heart attack and other cardiovascular diseases), and injuries (including road traffic, drowning, falls, fires, homicides, suicides, and other injuries). These COD clusters encompass 11 of the 34 PHMRC-defined COD categories. Figure \ref{fig:symptom_mean} shows that the prevalence of symptoms varies significantly with age.  For example, across all causes (seen in the bottom row of the figure) the prevalence of the interviewee reporting the decedent experienced weight loss increases with decedent age. However, for AIDS/TB deaths the prevalence of reported weight loss remains relatively stable across ages, and is much higher than in the full population. Estimates are more precise for CODs having more observations. 

Figure \ref{fig:symptom_cov} illustrates that the association between the set of example symptoms varies with both age group and COD.  Across all CODs (seen in the bottom row of the figure), the association between symptoms weakens with age. Using an extension of McNemar's test \citep{zhao2014testing}, the null hypothesis of homogeneous age group effects is rejected at the $0.05$ level for all $10$ of these illustrative symptom pairs (and for $72\%$ of the 14432 testable symptom pairs in the data set, i.e. those pairs of symptoms for which at least two of the category-level $2\times2$ tables have all nonzero margins-- results not shown, but code available in the online supplementary materials). This finding could be due to the fact that older people tend to have more ailments in general (imagine asking a grandmother about her everyday symptoms), while younger people likely experience clusters of symptoms relating to a specific illness. Across all ages (seen in the far right column of the figure), there are differences in associations between symptoms by cause. Again using the test of \citet{zhao2014testing}, the null hypothesis of homogeneous COD effects is rejected at the $0.05$ level for all $10$ of these illustrative symptom pairs (and for $88\%$ of all 14065 testable pairs in the data set). In summary, the data presented in Figures \ref{fig:symptom_mean} and \ref{fig:symptom_cov} show that both symptom prevalence and pairwise association vary by decedent age.

To the authors' knowledge, no existing methods consider potential covariate impacts on the prevalence or association of symptoms. Furthermore, few move beyond the conditional independence assumption (that is, that the probability of observing symptom are independent given COD). Only two current works explicitly model the conditional covariance between symptoms: \citet{li2018using} and \citet{kunihama2018bayesian}.The former models one common covariance structure across causes, while the latter models cause-specific symptom-level association via a latent factor model. The FActor Regression for Verbal Autopsy (FARVA) model described in the current paper has the most in common with \citet{kunihama2018bayesian}, which also models symptom-level association via a latent factor model. FARVA extends \citet{kunihama2018bayesian} in several key ways. It defines both the mean and the covariance of the latent variable associated with the symptom vector hierarchically so as to share information while still allowing flexibility across disparate causes. Furthermore, it allows the mean occurrence of symptoms and covariance between symptoms to depend on individual-level covariates (e.g., age). Section \ref{previous_work} discusses existing methods and distinct advantages of FARVA relative to the model of \citet{kunihama2018bayesian}. Sections \ref{prop_model} and \ref{sampler} describe the FARVA model and sampler. Section \ref{sim} describes a simulation experiment designed to separate aspects of performance of existing VA algorithms, and the model testing performed using real data. Section \ref{results} reports performance of select methods using simulated and PHMRC data with regard to CSMF accuracy and top cause assignment accuracy. Section \ref{disc} discusses future directions for this model and for the field of probabilistic COD assignment using VA data.

\section{Model}\label{model}

Let $y_i \in \{1,\ldots,C\}$ denote the underlying cause of death of person $i, i=1,\ldots,N$ and $\bm{s}_i = (s_{i1},\ldots,s_{iP})'$ be a vector of symptoms as measured by a verbal autopsy instrument. Consider the model:
\begin{equation}
\displaystyle \pi(y_i=c| \bm{s}_i) = \frac{\pi(\bm{s}_i | y_i=c)\pi(y_i=c)}{\Sigma_{c^{*}=1}^C \pi(\bm{s}_i | y_i=c^{*})\pi(y_i=c^{*})}.
\end{equation} 
Here the probability that the COD of person $i$ equals $c$ is modeled conditional on the reported symptoms. The predictive goal of the model is to infer COD in cases where only symptoms are observed and the true cause is unknown. Using the above model, one can also calculate the posterior cause specific mortality fraction (CSMF, the population proportion of deaths attributed to each cause; see equation (\ref{popdist}) in the Appendix) for the population of interest. The modeling choices remaining to be made are in how to treat the prior distribution over causes $\pi(y_i)$ and how to model the likelihood of symptoms given causes $\pi(\bm{s}_i | y_i).$

\subsection{Previous work}\label{previous_work}

Often in regions in which VA methods are adopted, there is little knowledge about the population CSMF. Even in cases where clinical cases in the living are studied, causes of disease are difficult to pin down \citep{crump2013etiology}. Furthermore, it is not necessarily known which diseases are most contributory to deaths, especially outside of the hospital setting. Thus the major distinguishing feature in the existing body of work on VA algorithms is in the treatment of $\pi(\bm{s}_i | y_i),$ rather than that of $\pi(y_i)$. 

Three commonly cited non-probabilistic methods are the Tariff method \citep{james2011performance}, the King-Lu (KL) method \citep{king2008verbal}, and the Simplified Symptom Pattern (SSP) method \citep{murray2011simplified}. The Tariff method is a score-based system that uses a heuristic (that tends to do well) rather than a probabilistic formulation. Tariff gives a score  to each combination of cause and symptom. These scores are used to assign a value to each possible death-cause combination, and this value is then used to rank CODs for each individual. An improved version of the Tariff method \citep{serina2015improving} is the foundation for the SmartVA-Analyze Application, available at \url{http://www.healthdata.org/verbal-autopsy/tools}. The KL method is designed to only estimate the CSMF and uses the assumption that the probability of symptoms given causes is the same in both the set of data for which COD is known (usually referred to as the training set) and that for which COD is unknown; it cannot infer individual cause of deaths (CODs). The closer this assumption is to the truth, and the more training data are available, the better this method will do. The SSP method uses the CSMF calculated from the KL method and calculates individual CODs using averages for the probability of symptoms given causes calculated across multiple random draws of symptoms. In each non-probabilistic method, some set of decedents having known COD are required for the model to learn a mapping between patterns of symptoms and COD.

InterVA was the first VA model to frame the relationship between symptoms and causes in a more probabilistic light \citep{byass2012strengthening}. Rather than relying on some training data for which COD is known, InterVA relies on a matrix of physician-generated scores assigned to the probability of observing each symptom given each cause (denoted by the $\text{P}(s|c)$ matrix, referring to the probability of symptom s given cause c). For example, physicians were asked to rate the probability that someone would answer ``Yes'' to ``Did the patient have a cough'' given that they died of TB, stroke, a motorcycle accident, etc. This matrix is useful because it allows models to be run in settings in which no training data are available. It is problematic because the values in this $\text{P}(s|c)$ matrix are likely not generalizable to all settings, are difficult to elicit (much of physician experience is likely in clinical cases and not deaths, and often the cause is not known), or may be internally inconsistent (because there are so many entries to fill out, it is very easy to, e.g., rank the probability of a patient having a cough for at least a month as higher than the probability of the patient having a cough for at least a week, even though the former case is a subset of the latter). InterVA models the probability of symptoms given causes as conditionally independent for all symptoms and causes.

There are three main issues with InterVA. First, the score used by InterVA to assign causes is based only on the presence of symptoms; it disregards symptoms when they are absent. Second, it is not possible to quantify uncertainty because the model does not contain features that are allowed to vary probabilistically. Third, InterVA is unable to incorporate physician-coded VA cases (i.e., gold standard data) into its algorithm. InSilicoVA (McCormick, 2016) is an extension of InterVA in both name and spirit. It addresses the three main issues with InterVA and provides a valid probabilistic framework in which the main ideas of InterVA can live. That is, InSilicoVA returns valid uncertainty estimates for both the COD and CSMF values. However, it still relies on the conditional independence assumption of InterVA. An even simpler framing for VA modeling directly uses a naive Bayes classifier (NBC) to assign the probability of a death being due to a given cause for each individual and each cause \citep{miasnikof2015naive}; the COD with the highest probability is then treated as the cause. Again, this framing relies on the conditional independence assumption.

Two more recent models that have relaxed the conditional independence assumption are the latent Gaussian graphical model of \citet{li2018using} and the factor model of \citet{kunihama2018bayesian}. Both models allow the prevalence of symptoms to vary by cause. The model of \citet{li2018using} assumes a shared covariance matrix between symptoms across all causes. On the other hand, the model of \citet{kunihama2018bayesian} defines separate covariance matrices for each cause. The former has the advantage of allowing for mixed data (i.e., not just binary indicators), but neither model allows for demographic, spatial, or temporal information to inform the prevalence of symptoms nor their covariance. 

\subsection{Proposed model}\label{prop_model}

\begin{figure}
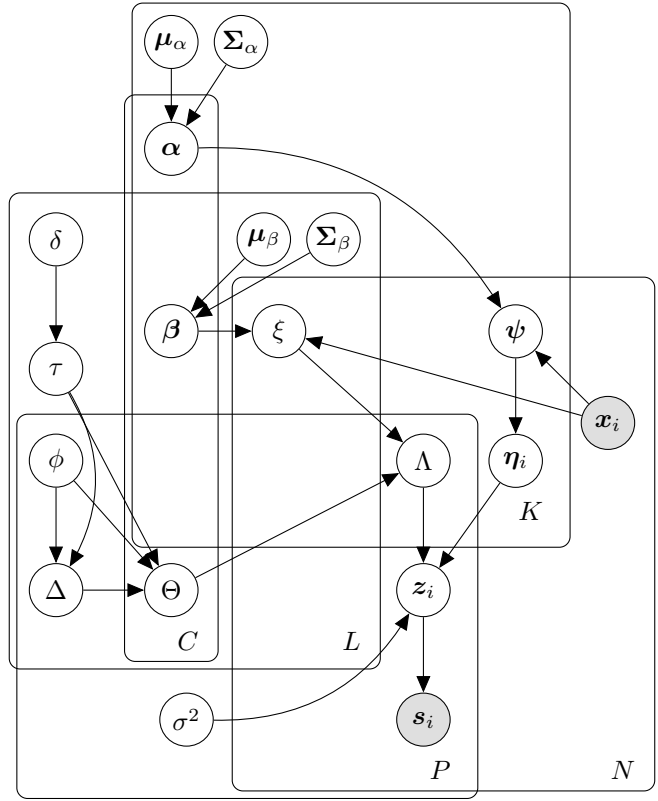

	\centering
	 \tikz{ %
		\node[obs] (s) {$\bm{s}_{i}$} ; %
		\node[latent, above=of s] (z) {$\bm{z}_{i}$} ; %
		\node[latent, left=of s, xshift=-1.4cm] (sig) {$\sigma^2$} ; %
		\node[latent, above=of z, yshift=0.0cm] (lambda) {$\Lambda$} ; %
		\node[latent, above right=of z, yshift=0.5cm] (eta) {$\bm{\eta}_i$} ; %
		\node[latent, above=of lambda, xshift=-1.9cm] (xi) {$\xi$} ; %
		\node[latent, above=of eta] (psi) {$\bm{\psi}$} ; %
		\node[obs, below right=of psi] (x) {$\bm{x}_{i}$} ; %
		\node[latent, left=of xi, xshift=0.3cm] (beta) {$\bm{\beta}$} ; %
		\node[latent, above=of beta, yshift=0.7cm] (alpha) {$\bm{\alpha}$} ; %
		\node[latent, above=of alpha, yshift=-0.3cm] (mualph) {$\bm{\mu}_{\alpha}$} ; %
		\node[latent, right=of mualph, xshift=-0.8cm] (sigalph) {$\bm{\Sigma}_{\alpha}$} ; %
		\node[latent, above=of xi, xshift=-0.2cm, yshift=-0.5cm] (mubet) {$\bm{\mu}_{\beta}$} ; %
		\node[latent, right=of mubet, xshift=-0.8cm] (sigbet) {$\bm{\Sigma}_{\beta}$} ; %
		\node[latent, left=of z, xshift=-1.6cm] (theta) {$\Theta$} ; %
		\node[latent, left=of theta, xshift=0.2cm] (delta) {$\Delta$} ; %
		\node[latent, above=of delta, yshift=-0.0cm] (phithet) {$\phi$} ; %
		\node[latent, above=of phithet, yshift=-0.5cm] (tauthet) {$\tau$} ; %
		\node[latent, above=of tauthet, yshift=-0.0cm] (delthet) {$\delta$} ; %
		
		\plate[inner sep=0.25cm, xshift=0.0cm, yshift=0.1cm] {plateC} {(beta) (alpha) (theta)} {$C$}; %
		\plate[inner sep=0.25cm, xshift=0.1cm, yshift=-0.1cm] {plateK} {(beta) (alpha) (xi) (psi) (lambda) (eta) (mualph) (sigalph)} {$K$}; %
		\plate[inner sep=0.25cm, xshift=0.0cm, yshift=0.1cm] {plateN} {(z) (s) (lambda) (eta) (xi) (psi) (x)} {$N$}; %
		\plate[inner sep=0.25cm, xshift=0.1cm, yshift=0.0cm] {plateP} {(theta) (z) (lambda) (s) (sig) (phithet) (delta)} {$P$}; %
		\plate[inner sep=0.25cm, xshift=-0.0cm, yshift=0.0cm] {plateL} {(delta) (beta) (xi) (phithet) (tauthet) (delta) (delthet) (mubet) (sigbet)} {$L$}; %
		
		\edge {mubet} {beta} ; %
		\edge {sigbet} {beta} ; %
		\edge {mualph} {alpha} ; %
		\edge {sigalph} {alpha} ; %
		\edge {delthet} {tauthet} ; %
		\edge {phithet} {delta} ; %
		\path (tauthet) edge[bend left, ->]  (delta) ;
		\edge {phithet} {theta} ; %
		\edge {tauthet} {theta} ; %
		\edge {delta} {theta} ; %
		\edge {theta} {lambda} ; %
		\path (alpha) edge[bend left, ->]  (psi) ;
		\edge {beta} {xi} ; %
		\edge {x} {xi, psi} ; %
		\edge {xi} {lambda} ; %
		\edge {psi} {eta} ; %
		\edge {lambda, eta} {z} ; %
		\path (sig) edge[bend right, ->]  (z) ;
		\edge {z} {s} ; %
	}
	\caption{Graphical representation of the FARVA model. Nodes shaded in gray are data (observed), unshaded nodes are parameters (learned by the model). Note $B,$ the number of covariates, is not shown but is the dimension of vectors $\bm{x}_i, \bm{\alpha}_{ck}, \bm{\mu}_{\alpha,k}, \bm{\beta}_{clk}, \bm{\mu}_{\beta,lk}$ and square matrices $\Sigma_{\alpha,k}, \Sigma_{\beta,lk}$.}
	\label{fig:graph_model}
\end{figure}

FARVA is a probabilistic model for VA data in which some decedents have known COD (i.e., there exists some labeled training data) and there is interest in learning about individual COD, population CSMF, and/or the mean and covariance structure of responses in the symptom questionnaire. The explicit goals of the FARVA model proposed here are to capture dependence of symptoms given a cause; share information across causes via hierarchical modeling to improve estimates associated with causes having few observed deaths; allow both the conditional prevalence and the conditional association between symptoms to vary with covariates (e.g., age of patient, time of year, geographic region); probabilistically predict cause of death for a new individual given their symptoms; and improve on COD and CSMF estimation relative to current state-of-the-art VA algorithms. A graphical representation of the model is shown in Figure \ref{fig:graph_model}, the parameters are summarized in Tables \ref{tab:model_terms1} and \ref{tab:model_terms2}, and a toy example to explain how the model works in lower dimensions is described in the online supplementary materials.

\subsubsection{Prior over causes}\label{prior}

Using a Bayesian approach to characterize uncertainty in the proportions of deaths in each COD category, let:
\begin{equation} 
\{\Pr(y_i=1), \ldots, \Pr(y_i=C)\} \sim \text{Dirichlet}(a_1,...,a_C).
\end{equation}
Under the assumption that little is known about the CSMF in the region of interest and the number of possible causes is high relative to the number of common causes, set $a_1 = \ldots = a_C < 1.$ This choice leads to equal probability of each cause a priori (i.e., $\pi(y_i=c) \propto 1 \ \forall \ c \in \{1,\ldots,C\}$ where $\pi(y_i)$ is a probability distribution over causes of death) and encourages concentration near sparse subvectors. This assumption is likely valid in real applications; e.g., the gold standard PHMRC data set exhibits many causes having few observed deaths (see Figure \ref{fig:phmrc_causes}). However, this prior can easily be modified to incorporate prior knowledge about the CSMF or to encourage uniformity in the CSMF. For example, if a pilot study performed in the region of interest found counts $n_c$ attributable to COD $c, c=1\ldots C$, then an option for an informed prior is $a_1 = n_1, \ldots = a_C = n_C.$ If uniformity was desired, one could set $a_1 = \ldots = a_C = 1.$


\subsubsection{Likelihood of symptoms given a cause}\label{likelihood}

Recall that the likelihood takes the form $\pi(\bm{s}_i | y_i),$ i.e. the distribution of the symptom vector $\bm{s}_i $ is conditioned on COD $y_i$. Throughout the likelihood description, the motivation behind each modeling decision and a short description of said decision will be stated in bold preceding each mathematical description.

\textbf{Motivation: Symptom data includes questions on duration of symptoms in addition to their presence/absence. Therefore, allow the model to encompass data of mixed type.} In order to allow the FARVA model to encompass data of binary, continuous, count, and categorical type, in the specification of $\pi(\bm{s}_i | y_i)$ define:
\begin{align}
s_{ij} &= f_j(z_{ij}), \quad  j=1,\ldots,P,  
\end{align}
where $\bm{z}_i=(z_{i,1},\ldots,z_{iP})'$ is a vector of latent continuous symptoms. The particular link function $f_j$ depends on the symptom specification, allowing for mixed scale data via selection of an appropriate link by scale and type. Let $f_j(z_{ij}) = z_{ij}$ or $f_j(z_{ij}) = \text{log}(z_{ij})$ for continuous $s_{ij}$, with the latter chosen for strictly positive and positively skewed cases. Scale continuous symptoms (after a log transformation if applicable) to have unit variance. Let $f_j(z_{ij}) = 1(z_{ij}>0)$ for binary $s_{ij},$ where $1(\cdot)$ is an indicator function taking the value of 1 when the argument is true and 0 when the argument is false. Finally, for (potentially zero-inflated) count $s_{ij}$ let $f_j(z_{ij})$ be a rounding operator such that $f_j(z_{ij}) = 0$ if $z_{ij}<0$ and $f_j(z_{ij}) = k$ if $k-1 \leq z_{ij} < k$ \citep{canale2013nonparametric}. Categorical variables may be handled by choosing one category as the baseline and transforming the $T$ categories into $T-1$ binary variables indicating whether or not the categorical value for that individual took that non-baseline category value (so either none or one of the $T-1$ variables will take on a value of 1).  Other specifications of $f$ would be needed for ordered categorical data to fit within this framework, if desired.  Note that of existing models, only \citet{li2018using} allows for non-binary data, specifically continuous data.

\textbf{Motivation: It is likely that the full set of symptoms are observations of some lower dimensional syndromic state. Thus, model symptom mean and covariance structure via a smaller set of underlying (unobserved) factors.} The number of measured symptoms, $P$, may be quite large for verbal autopsies (e.g., the 2016 WHO VA instrument \citep{va2016} asks up to 253 questions for adult and child deaths). Furthermore, many of the causes of interest may have very few observed deaths (e.g., see deaths by cause in the gold standard PHMRC data set in Figure \ref{fig:phmrc_causes}). Rather than estimating $P(P+1)/2$ parameters for each of the $C$ conditional covariance matrices, FARVA introduces a novel hierarchical latent factor model for the latent continuous vector $\bm{z}_i$ underlying $\bm{s}_i.$ To begin, define:
\begin{align}\label{eq:z_i_1}
\bm{z}_i &= \Lambda_{c[i]}(\bm{x}_i)\bm{\eta}_i + \bm{\epsilon}_i, \quad \bm{\eta}_i \sim \text{N}(\bm{\psi}_{c[i]}(\bm{x}_i),I_K),  \quad \bm{\epsilon}_i \sim \text{N}(\bm{0}_P, \Sigma_0),
\end{align}
where $c[i]$ indexes the COD for person $i$, $\bm{x}_i$ are predictors (i.e. covariates) associated with person $i$, and $K$ is the number of latent factors; in practice $K$ is allowed to be unknown. Here, $\Lambda_{c[i]}(\bm{x}_i)$ is a predictor-dependent $P \times K$ factor loadings matrix, $\bm{\eta}_i$ is a predictor-dependent $K \times 1$ latent factor vector, and $\bm{\epsilon}_i$ is the $P \times 1$ independent noise vector. The noise covariance $\Sigma_0$ is a diagonal matrix having entries $\sigma_j^2, j=1\ldots P,$ with $\sigma_j^2$ fixed to equal 1 when $s_{ij}$ is binary or categorical and estimated otherwise. In the above model, $\bm{\eta}_i$ captures a set of latent syndromes, $\bm{\psi}_{c[i]}(\bm{x}_i)$ the covariate-dependent mean of these latent symptoms for COD $c[i]$, and $\Lambda_{c[i]}(\bm{x}_i)$ is a low-rank description of the symptom correlations for COD $c[i]$ with covariates $\bm{x}_i.$ Note that throughout the paper the convention $\text{N}(\text{mean}, \text{variance/covariance})$ is used in parameterizing the normally distributed terms.

Since $\bm{\eta}_i$ is not actually known, dependence is induced in the latent $z_i$ via the marginalization over $\bm{\eta}_i.$ This induced dependence in elements of $z_i$ creates the possibility for dependence to be captured in the likelihood of symptoms given a cause. Specifically, the prior induced on the latent $z_i$ by integrating out $\bm{\eta}_i$ is:
\begin{equation}
\bm{z}_i|y_i \sim \text{N}(\Lambda_{c[i]}(\bm{x}_i)\bm{\psi}_{c[i]}(\bm{x}_i), \Lambda_{c[i]}(\bm{x}_i) \Lambda_{c[i]}(\bm{x}_i)' + \Sigma_0),
\end{equation}
leading to a parsimonious representation of both the mean and the dependence between the $P$ symptoms. An alternative is to define the model as $\bm{z}_i = \bm{\mu}_{c[i]}(\bm{x}_i) + \Lambda_{c[i]}(\bm{x}_i)\bm{\eta}_i + \bm{\epsilon}_i$, with $\bm{\eta}_i \sim \text{N}(\bm{0}_K,I_K)$, i.e. with a $P$-dimensional mean regression. This direct parameterization of the mean has the advantage of allowing a more direct/interpretable way to encode prior information about conditional symptom prevalence, but in the case of large $P$ it is more computationally intensive.

\textbf{Motivation: The association between symptoms likely depends on factors such as season and decedent age, but modeling a number of covariate-dependent parameters whose size depends on the number of symptoms is computationally intensive. Define a computationally efficient structure by which to model covariate-dependent covariance structure.} In order to estimate covariate-dependent factor loadings for large $P$, decompose $\Lambda_{c[i]}(\bm{x}_i)$ to be a weighted combination of a smaller set of basis elements, as in \citet{fox2015bayesian}. Namely, let 
\begin{align}\label{eq:z_i_2}
\begin{split}
\Lambda_{c[i]}&(\bm{x}_i) = \Theta_{c[i]} \bm{\xi}_{c[i]}(\bm{x}_i), \\ 
&\Theta_{c[i]} \in \mathbb{R}^{p \times L}, \quad \bm{\xi}_{c[i]}(\bm{x}_i) = \{\xi_{c[i],lk}(\bm{x}_i), l=1,\ldots,L, \ k=1,\ldots,K\},
\end{split}
\end{align}
where $\Theta_{c[i]}$ is the coefficient matrix with weights mapping the smaller set of predictor-dependent basis functions $\bm{\xi}_{c[i]}(\bm{x}_i)$ to the higher-dimensional loadings matrix $\Lambda_{c[i]}(\bm{x}_i).$ The number of covariate-dependent parameters to be estimated is $P(P+1)/2$ in direct modeling of the conditional covariance matrix, $KP$ in (\ref{eq:z_i_1}), and $KL$ in (\ref{eq:z_i_2}). Here, $K$ is the dimension of the subspace that is assumed to capture the statistical variability in $\bm{z}$ and $L$ is the maximum size of the basis for any given value of $K$. Typically $K\ll P$ and $L \ll P,$ yielding significant computational savings by using the formulation in (\ref{eq:z_i_2}). If the dependence structure underlying the relationship between symptoms were actually highly complex it would be expected that $K$ and $L$ would need to approach $P$ in size, negating computational advantages of the factor model framework.

\textbf{Motivation: The size of the underlying basis defining relationships between symptoms isn't actually known. Furthermore, these relationships may be difficult to estimate when the number of observed deaths of a given cause is small. Thus, learn (rather than fix) the basis size while sharing information across causes hierarchically.} To share information across causes define the entries of each coefficient matrix $\Theta_{c}, c=1,\ldots, C,$ to share a common population level mean, $\Delta,$ across causes. This hierarchical structure serves the purpose of linking factor loadings and increasing robustness when estimating the factor loadings for uncommon causes. Thus similar causes should have similar factor loadings, and loading estimates for rare causes will tend toward the overall population-level mean loadings. Sparsity is induced on the population mean parameter for each entry in the coefficient matrix via the adaptive shrinkage prior of \citet{bhattacharya2011sparse}. The hierarchical shrinkage prior on each element of $\Theta_{c[i]}$ is shown below:
\begin{align}
\begin{split}
\theta_{c[i], jl} \sim & \text{N}(\Delta_{jl}, \phi_{jl}^{-1} \tau_{l}^{-1}), \\ 
\Delta_{jl} \sim & \text{N}(0, \phi_{jl}^{-1} \tau_{l}^{-1}), \quad  \phi_{jl} \sim \text{Ga}(\gamma/2,\gamma/2), \quad  \tau_{l}=\prod_{h=1}^l \delta_{h}, \\ 
&  j=1,\ldots,P, l=1,\ldots,L,
\end{split}
\end{align} 
where $L$ is a conservative upper bound on the basis dimension. The shrinkage prior of \citet{bhattacharya2011sparse} with specification suggested by \citet{durante2017note} is used on the entries of both the within-cause coefficient matrix $\Theta$ and across-causes mean matrix $\Delta.$ This prior specifies $\phi_{jl}$ to be a local precision specific to element $j,l$ and $\tau_{l}$ to be a column-specific multiplier. Specifically, this shrinkage prior pulls entries in later columns more strongly toward the mean by letting $\delta_{1} \sim \text{Ga}(d_{1},1)$ and $\delta_{h} \sim \text{Ga}(d_{2},1), h \geq 2,$ with $d_{2} > 1.$  The aim of this prior is to increase the degree of shrinkage as the column index of the matrix grows. In the case of $\Theta,$ cause-specific entries are shrunk towards the population mean $\Delta$. In the case of $\Delta,$ the population mean is shrunk towards zero. Because the precision terms $\phi_{jl}$ and $\tau_{l}$ are shared across $\theta_{c[i], jl}$ and $\Delta_{jl}$, the overall effect is shrinkage of columns of $\Theta$ toward columns of $\Delta$, which approach 0. The result of this shrinkage is an effective truncation of the number of latent factors because the corresponding rows of the basis $\bm{\xi}$ will then have insignificant effect in the definition of the resulting mean and covariance of $\bm{z}$.  A hallmark of $L$ not being large enough would be columns of $\Delta$ close to the $L$th column remaining large in spite of the shrinkage prior on its entries.

\textbf{Motivation: The association between symptoms likely depends on covariates, but this covariate dependent relationship may be difficult to learn for causes with a small number of observed deaths. Therefore, model covariate-dependent covariance structure while sharing information across causes hierarchically.} The choice of how to model each $\bm{\xi}_{c[i]}(\bm{x}_i),$ the cause-specific $L \times K$ matrix of basis functions of equation (\ref{eq:z_i_2}), to include covariates is flexible. A non-hierarchical nonparametric model can be introduced by choosing independent Gaussian process priors on entries of $\bm{\xi},$ i.e. $\xi_{c[i],lk} \sim \text{GP}(0,\sigma_{\xi}(\cdot))$ with $\sigma_{\xi}(\bm{x},\bm{x}')$ an appropriate kernel function. If the covariate included is spatial location, for example $\bm{x}=(lat, lon)'$, a squared exponential covariance kernel with unit variance could be used (the restriction that the kernel has unit variance arises due to identifiability issues with the multiplication with $\Theta$). A separable covariance structure across space and time could also be chosen. The advantage of such a formulation is the ability to model symptom dependence that varies smoothly across space and time, and to infer dependence while predicting COD at locations for which no VAs were recorded. Parametric models for each entry could also be considered by letting $\xi_{c[i],lk}(\bm{x}_i) = \bm{\beta}_{c[i],lk}^T \bm{x}_i.$ Here $\bm{\beta}$ is a $B \times 1$ vector, where $B$ is the number of covariates included in the model. This form would be suitable for, e.g., including an indicator variable for whether the $i$th person died in the hospital, whether the death occured during the malaria season, or some other relevant information. Aspects of the interviewer, e.g., level of experience, could also be included as predictors. Smoothing splines, piecewise splines, or combinations of parametric and nonparametric forms for predictors can be included, but the level of complexity it is possible and useful to include will be limited by the expressiveness and dimension of the data.

In order to borrow information in the case of small sample sizes within a given cause, introduce a hierarchical prior on elements of the predictor-dependent basis functions $\bm{\xi}$ and parametric regression parameters $\bm{\beta}$. Rather than modeling $\bm{\beta}_{c[i],lk}$ independently for each $c$, model $\bm{\beta}_{c[i],lk}$ as coming from a shared population mean parameter $\bm{\mu}_{\beta_{lk}}.$ For convenience of notation, focus on the case when the covariates of interest are the same for the mean and covariance components of the model. Define the following prior on elements $\xi_{c[i],lk}$ of $\bm{\xi}_{c[i]}:$
\begin{align}\label{eq:xi_regression}
\begin{split}
\xi_{c[i],lk}(\bm{x}_i) = &\bm{\beta}_{c[i],lk}^T \bm{x}_i,  \\
& \bm{\beta}_{c[i],lk} \sim \text{N}_B(\bm{\mu}_{\beta_{lk}}, \Sigma_{\beta_{lk}}) ,  \\ 
& \bm{\mu}_{\beta_{lk}} \sim \text{N}_B(\mu_0, \Lambda_0), \quad \Sigma_{\beta_{lk}} \sim \text{IW}(\nu_0, S_0), \\
&  l=1,\ldots,L, k=1,\ldots,K.
\end{split}
\end{align}

\textbf{Motivation: Average symptom observations likely depend on covariates, but this covariate dependent relationship may be difficult to learn for causes with a small number of observed deaths. Model covariate-dependent mean structure while sharing information across causes hierarchically.} The choice of how to model $\bm{\eta}_{i}$, the cause-specific $K \times 1$ vector of latent syndromes of equation (\ref{eq:z_i_1}), to include covariate information is similarly flexible to that of how to model $\bm{\xi}$. As with the model for $\bm{\xi}_{c[i]}(\bm{x}_i),$ a non-hierarchical nonparametric model can be introduced by choosing independent Gaussian process priors on entries of $\bm{\psi}_{c[i]}.$ A parametric model is defined by fixing $\psi_{c[i],k} = \bm{\alpha}_{c[i],k}^T \bm{x}_i.$  This choice induces the following hierarchical formulation for elements $\eta_{i,k}$ of $\bm{\eta}_{i}$ :
\begin{align}\label{eq:eta_regression}
\begin{split}
\eta_{i,k}(\bm{x}_i) = &\psi_{c[i],k} + \epsilon_{\eta_{k}i} \\
= &\bm{\alpha}_{c[i],k}^T \bm{x}_i + \epsilon_{\eta_{k}i}, \quad \epsilon_{\eta_{k}i} \sim \text{N}(0,1) \\
& \bm{\alpha}_{c[i],k} \sim \text{N}_B(\bm{\mu}_{\alpha_{k}}, \Sigma_{\alpha_{k}}) ,  \\ 
& \bm{\mu}_{\alpha_{k}} \sim \text{N}_B(A_0, L_0), \quad \Sigma_{\alpha_{k}} \sim \text{IW}(v_0, D_0), \\
&  k=1,\ldots,K.
\end{split}
\end{align}

When there are no covariates, model elements of $\bm{\eta}_{i}$ and $\bm{\xi}_{c[i]}$ using a mean-only model. Finally, it is possible to include covariates in the model for $\bm{\xi}_{c[i]}(\cdot)$ that are not included in that for $\bm{\eta}_{i}(\cdot)$ or vice versa.


\begin{table}
	\footnotesize
	\bgroup
	\def\arraystretch{1.2}
	\begin{center}
		\begin{tabular}{||l l||} 
			\hline
			Variable & Description \\ [0.5ex] 
			\hline\hline
			$s_{ij}$ & Observed symptom $j$ of decedent $i$ \\ 
			$\bm{x}_{i}$ & Observed covariates for decedent $i$ \\ 
			$z_{ij}$ & Latent continuous symptom $j$ of decedent $i$ \\
			$\sigma_j^2$ & Variance for latent continuous symptom $j$ \\ 
			$\Lambda_{ijk}$ & $jk$-th entry of decedent $i$'s factor loadings matrix \\
			$\eta_{ik}$ & $k$-th entry of decedent $i$'s latent factor vector \\
			$\psi_{ik}$ & $k$-th entry of decedent $i$'s latent factor mean \\
			$\bm{\alpha}_{ck}$ & Regression coefficients for $k$-th entry of $\bm{\psi}$ for decedents dying of cause $c$ \\
			$\bm{\mu}_{\alpha,k}$ & Global mean of cause-specific regression coefficients for $k$-th entry of $\bm{\psi}$ \\
			$\Sigma_{\alpha,k}$ & Global covariance of cause-specific regression coefficients for $k$-th entry of $\bm{\psi}$ \\
			$\xi_{ilk}$ & $lk$-th entry of decedent $i$'s latent predictor-dependent basis functions \\
			$\bm{\beta}_{clk}$ & Regression coefficients for $lk$-th entry of $\xi$ for decedents dying of cause $c$ \\
			$\bm{\mu}_{\beta,lk}$ & Global mean of cause-specific regression coefficients for $lk$-th entry of $\xi$ \\
			$\Sigma_{\beta,lk}$ & Global covariance of cause-specific regression coefficients for $lk$-th entry of $\xi$ \\
			$\Theta_{cjl}$ & $jl$-th entry of coefficient matrix mapping $\bm{\xi}$ to $\Lambda$ for decedents dying of cause $c$ \\
			$\Delta_{jl}$ & $jl$-th entry of global mean of cause-specific coefficient matrices $\Theta_{c}$ \\
			$\phi_{jl}$ & Local precision specific to element $jl$ of the global and cause-specific coefficient matrices \\
			$\tau_{l}$ & Column-specific precision for $l$-th column of the global and cause-specific coefficient matrices \\
			$\delta_{l}$ & $l$-th component of the multiplicative gamma process forming column-specific precision $\tau_{l}$ \\ [0.5ex] 
			\hline
		\end{tabular}
	\end{center}
	\caption{Summary of the parameters of the FARVA model.}
	\label{tab:model_terms1}
	\egroup
\end{table}

\begin{table}
	\footnotesize
	\bgroup
	\def\arraystretch{1.2}
	\begin{center}
		\begin{tabular}{||l l||} 
			\hline
			Plate & Description \\ [0.5ex] 
			\hline\hline
			$N$ & Decedents $i=1,\ldots,N$ \\ 
			$P$ & Symptoms $j=1,\ldots,P$ \\ 
			$K$ & Latent factors $k=1,\ldots,K$ \\ 
			$C$ & Causes of death $c=1,\ldots,C$ \\ 
			$L$ & Maximum basis size $l=1,\ldots,L$ \\ 
			$B$ & Covariates $b=1,\ldots,B$ \\ [0.5ex] 
			\hline
		\end{tabular}
	\end{center}
	\caption{Summary of the indices (plates in Figure \ref{fig:graph_model}) of the FARVA model.}
	\label{tab:model_terms2}
	\egroup
\end{table}

\subsubsection{Model assumptions and limitations}

The FARVA model assumes the symptom means can be specified via the linear projection of latent factors into the symptom space ($\Lambda_{c[i]}(\bm{x}_i)\bm{\psi}_{c[i]}(\bm{x}_i)$). In a practical sense, this is analogous to assuming that the full set of symptoms arise from some smaller set of underlying lower-dimensional syndromes. If symptom means do not arise from  such a process, or if the number of latent factors required to capture symptom means well is large, it is preferable to simply use a separate mean vector as discussed in Section \ref{likelihood}.

The model also assumes the covariance between symptoms can be specified by a structured low-rank matrix ($\Lambda_{c[i]}(\bm{x}_i) \Lambda_{c[i]}(\bm{x}_i)'$) and a diagonal noise matrix ($\Sigma_0$). If the rank of the structured component needs to be large, i.e. as $K$ approaches $P$, the computational benefits of the factor model specification disappear and the amount of data required to learn all of the model parameters becomes impractically high for the VA setting.

Recall that $K$ need only be specified as a conservative upper bound on the number of latent factors believed to be necessary for modeling. If the chosen $K$ is too small (diagnosable by observing that later columns of $\Delta$ do not converge to 0) model performance and convergence may suffer. If the chosen $K$ is too large (diagnosable by observing that many columns of $\Delta$, not only the final few, have converged to 0) computation time will be higher than necessary but performance and convergence should not be impacted.

Modelers face decisions as to the inclusion/specification of (1) symptoms and (2) covariates. Regarding (1) one could, for example, either exclude or include information on coughing as a symptom. If the latter, one could decide to use a binary specification (e.g., did or didn't cough) or a numeric time specification (e.g., coughed for $x$ days before death). Regarding (2) one could, for example, either exclude or include age as a covariate. If the latter, one could use a binary specification (e.g., elder vs. non-elder) or a numeric year specification (e.g., 68 vs. 22 years old) in a linear model. As noted in Section \label{likelihood} it would also be possible to include certain covariates via nonlinear models (e.g., Gaussian processes).

A possible Bayesian solution to model selection would be to compute posterior model probabilities and Bayes factors for each symptom/covariate specification of interest. These quantities rely on the calculation of the marginal likelihood, an analytically intractable integral. Approximating the marginal likelihood for FARVA, e.g. via bridge sampling, could aid in understanding and formally deciding which symptoms and/or decedent characteristics are important when modeling CODs. A simpler alternative for exploring model predictive performance under different possible sets of symptoms/covariates is to perform cross validation, in which the model is trained/tested using each input set. 

\subsection{Posterior computation}\label{sampler}

The posterior for the FARVA model is not available in closed form, but it may be approximated via samples obtained from a Markov chain Monte Carlo (MCMC) algorithm. Closed form full conditional distributions of the parameters associated with the model allow the use of a straightforward Gibbs sampler for these draws. Full details on each Gibbs step, posterior computations, hyperparameter settings, and the practical considerations on how the steps are modified for real data are provided in the online supplementary materials.


\subsubsection{Performance metrics}\label{perf_met}

The performance metrics used to assess model accuracy are top cause accuracy $\text{ACC}_1$ (\ref{eq:acc}), a measure for how well CODs for individuals are predicted, and CSMF accuracy $\text{ACC}_{\text{CSMF}}$ (\ref{eq:acccsmf}), a measure for how well the population CSMF is predicted. The top cause accuracy and CSMF accuracy take possible values from 0 to 1, with 1 being best. 
\begin{equation}\label{eq:acc}
	\text{ACC}_1 = \frac{\text{\# of correct COD being the top cause assignment}}{N}
\end{equation}
\begin{equation}\label{eq:acccsmf}
\text{ACC}_{\text{CSMF}} = 1 - \frac{\sum_{c=1}^C |\text{CSMF}_c^{true} - \text{CSMF}_c^{pred}| }{2(1-\text{min}(\text{CSMF}^{true}))}
\end{equation}
The CSMF accuracy formula was defined in \citet{murray2011robust}, with the idea being that the worst case scenario for CSMF prediction is to put all predicted CSMF weight on the least common cause. This scenario corresponds to a total absolute error of $2(1-\text{min}(\text{CSMF}^{true}))$. Note that these metrics all rely on there being some hold-out data set for which information on COD is known but can be hidden from the model during training. Results for chance-corrected concordance, a metric with a 0 value corresponding to the performance expected of random uniform guessing, are included in the online supplementary materials.

\subsection{Simulations}\label{sim}

Commonly, work on VA based COD algorithms jump immediately to training/testing using samples from the PHMRC data set. This approach has the advantage of allowing questions such as ``How do algorithms trained in one context and tested in another perform?'' and ``What causes are easier/harder to predict?''; it places the emphasis on performance of algorithms with real data. However, the underlying mean/covariance of symptoms in these resampled PHMRC data sets are unknown. This limitation makes it difficult to manipulate and explore how differing symptom level mean and covariance structures impact model performance. To address this question, the current work includes simulated scaled down data sets having just 928 observations, 21 ``symptoms'', and 4 ``causes''. Via different generative processes, it is possible to explore various structures of cause-specificity and covariate dependence for the simulated datasets' mean and covariance.

Table \ref{sim_table} shows detailed information about the various simulation configurations. For each simulation configuration sample 1000 data sets, a number chosen to balance the considerations of Monte Carlo error and computation time, and run FARVA, Kunihama's Bayesian factor (BF) model, open source Tariff, InsilicoVA, and the naive Bayes classifier (NBC). Broadly speaking, there are two classes of configuration:
\begin{enumerate}
	\item No covariate information included: Simulations \textit{a} through \textit{d}, with various combinations of cause-specificity in mean/covariance. FARVA is likely to only do as well as existing methods.
	\item Covariate information included: Simulations \textit{e} through \textit{g}, with various combinations of cause-specificity in mean/covariance, with covariate effects. FARVA is expected to outperform existing methods.
\end{enumerate}
Note that although simulation \textit{g} includes mixed and continuous versions of the data set for testing with FARVA, all other methods compared can use only binary data. More details of how data are simulated and visuals of the impact of various simulation settings are provided in the online supplementary materials.

\begin{table}
	\footnotesize
	\centering
	\begin{tabular}{ l c||c|c|c  } 
		\toprule
		\multicolumn{2}{c||}{} &\multicolumn{3}{c}{\textbf{Artificial data structure}} \\
		\multicolumn{1}{c}{} & \multicolumn{1}{c||}{Simulation} & 
		\multicolumn{1}{c}{Mean structure} & 
		\multicolumn{1}{c}{Covariance structure} & 
		\multicolumn{1}{c}{Data type} \\
		\midrule
		\multirow{2}{*}{\begin{minipage}{2cm}No covariate information included\end{minipage}} & \textit{a}  & S & C, I &  B \\
		& \textit{b} &  S & C, D   & B \\
		& \textit{c} & C & S, D &  B \\
		& \textit{d} & S & S, D & B \\
		\midrule
		\multirow{2}{*}{\begin{minipage}{1.8cm}Covariate information included\end{minipage}} & \textit{e} & S, V & C, I & B \\
		& \textit{f} & C & S, D, V & B  \\
		& \textit{g}$_1$ & S, V & S, D, V & B \\
		& \textit{g}$_2$ & S, V & S, D, V & M \\
		& \textit{g}$_3$ & S, V & S, D, V & T \\
		\bottomrule
	\end{tabular}
	\caption{\label{sim_table} In the table above S denotes cause-specific, C denotes common across causes; I denotes independent, D denoted non-independent; V denotes covariate-dependent; B denotes binary, M denotes mixed, and T denotes continuous. For example, simulation $d$ has cause-specific (S) mean structure and cause-specific (S) dependent (D) covariance structure, and the data type is binary (B). The simulation letter denotes that 1000 data sets are sampled for that letter. For example, simulations $g_1,$ $g_2,$ and $g_3$ are based on the same underlying data set so the same inputs are provided for each to the BF, InSilicoVA, NBC, and Tariff methods, but different inputs are given to FARVA ($g_1$ uses binary inputs, $g_2$ uses mixed inputs, and $g_3$ uses continuous inputs).} 
\end{table}

Simulations using the PHMRC data are also performed. To begin, the data are read into \textbf{R} using the command \texttt{read.csv(openVA::getPHMRC\_url("adult"))} (yielding 7841 observations) and the one observation for which a decedent has age $<12$ is removed (leaving 7840 adult deaths). The data are then divided by site (Andhra Pradesh, India ; Bohol, Philippines; Dar es Salaam, Tanzania; Mexico City, Mexico; Pemba Island, Tanzania; and Uttar Pradesh, India). For each site the following is repeated 100 times: First, site-specific data are split into 75\% training, 25\% test. Then, data cleaning steps used in the \textbf{OpenVA} software were performed, i.e. all variables converted to dichotomous symptoms matching those used in InterVA algorithm. Specifically, the adult PHMRC data was processed using the \texttt{ConvertData.phmrc()} command with \texttt{cutoff = "default"} and \texttt{cause = "va34"}. Variables were then cleaned such that symptoms having a  high missingness rate (specifically, over 95\% missing) or zero variability (i.e., all respondents had the same answer) were removed from analysis. Only the subset of causes included in the training data set were included as possible causes in each analysis. Finally, each model is run, with FARVA including whether or not each decedent was an elder ($\geq 65$) as a covariate. The competitor methods included are the same as those used in the simulated data runs. The InterVA and King Lu methods were not included as competitors, as both were found by \citet{kunihama2018bayesian} to have consistently poor performance relative to state of the art.  

 To explore cause-specific and potentially covariate-dependent prevalence and patterns of association between symptoms, the FARVA model is run using the full PHMRC data set (i.e., including data from across all sites and with no held-out test data) with (1) no covariates included and (2) a covariate for whether or not each decedent was an elder included. This exploration is a novel contribution to the VA literature. \citet{li2018using} described overall patterns of association between symptoms, but their model does not allow for covariance structure to differ across causes. No previous paper has explored the difference in patterns of association between symptoms across CODs (although the BF model would also allow such an analysis, none was performed in \citet{kunihama2018bayesian}). Furthermore, as no previous models have allowed for covariate inclusion, the question of how the mean or covariance structure differs within (or across) COD(s) for a given covariate has yet to be addressed for any covariate. Note that results here are meant to be illustrative rather than exhaustive, highlighting the utility of the FARVA model for exploring both the symptom level mean structure and patterns of association. Further simulation run details are provided in the online supplementary materials.

\section{Results and discussion}\label{results}

Performance of both the simulated and the PHMRC data runs are discussed below. In summary, explicitly accounting for covariate dependence in the model does improve predictive performance in both simulated and real data. The degree to which performance improves depends on the strength of the relationship between a covariate and symptom prevalence and/or associations.

\subsection{Simulation performance}

Two sets of simulations are considered. In the first, no covariate information is included or modeled (simulations $a$ through $d$). In the second, a single binary covariate is introduced as a modifier of the mean and/or covariance structure of symptoms (simulations $e$ through $g_3$). See the online supplementary materials for visual examples of the symptom mean and covariance structure from each simulation setting, and for performance as a function of degree of cause-specificity, and for plots of performance as a function of covariate dependence and proportion of data allowed to be continuous.

\subsubsection{Simulations with no covariate dependence}

Simulation \textit{a} fixes cause-specific symptom mean structure for each cause and shared \textbf{independent} covariance structure across causes. This simulation is designed to mimic a scenario in which the conditionally independent assumption is valid, and the only information about cause of death comes from symptom prevalence. The models allowing the specification of a non-independent conditional covariance matrix (FARVA and BF) perform comparably to models assuming conditional independence. Simulation \textit{b} is similar to simulation \textit{a} in that the only information on COD is found in the differing mean structure across causes. The difference is that symptoms are assumed to have a shared \textbf{dependent} covariance structure across causes. Both FARVA and BF are able to learn this shared covariance structure, and both perform slightly better than models assuming conditional independence.

\begin{table}
	\footnotesize
	\centering
	\begin{tabular}{ cc||ccccc  }
		\toprule
		\multicolumn{2}{c||}{} &\multicolumn{5}{c}{\textbf{VA model}} \\
		\multicolumn{1}{c}{} & \multicolumn{1}{c||}{Simulation} & 
		\multicolumn{1}{c}{FARVA} & 
		\multicolumn{1}{c}{BF} & 
		\multicolumn{1}{c}{InSilico} & 
		\multicolumn{1}{c}{NBC} & 
		\multicolumn{1}{c}{Tariff} \\
		\midrule
		\multirow{2}{*}{\begin{minipage}{1cm}$\text{ACC}_1$\end{minipage}} & \textit{a}   & 0.524 (0.047) & 0.518 (0.047) & 0.493 (0.069) & \textbf{0.528} (0.046) & 0.50 (0.045) \\
		& {b} &  \textbf{0.532} (0.045) & 0.527 (0.045) & 0.470 (0.068) & 0.522 (0.048) & 0.498 (0.045) \\
		& \textit{c} & \textbf{0.635} (0.039) & 0.634 (0.040) & 0.268 (0.065) & 0.275 (0.035) & 0.262 (0.040) \\
		& \textit{d} & 0.711 (0.036) & \textbf{0.712} (0.037) & 0.420 (0.070) & 0.468 (0.042) & 0.444 (0.043) \\
		\midrule
		\multirow{2}{*}{\begin{minipage}{1cm}$\text{ACC}_\text{CSMF}$\end{minipage}} & \textit{a}   & 0.894 (0.013) & 0.895 (0.014) & 0.841 (0.104) & \textbf{0.908} (0.024) & 0.903 (0.027) \\
		& \textit{b} & 0.895 (0.014) & 0.898 (0.016) & 0.831 (0.104) & 0.901 (0.034) & \textbf{0.902 (0.030)} \\ 
		& \textit{c} & 0.913 (0.014) & \textbf{0.915} (0.015) & 0.650 (0.148) & 0.858 (0.040) & 0.808 (0.087) \\
		& \textit{d} & 0.926 (0.015) & \textbf{0.929 (0.016)} & 0.784 (0.139) & 0.896 (0.030) & 0.892 (0.031) \\
		\bottomrule
	\end{tabular}
	\caption{\label{sim_table_res_1} Above are the mean (SD) of the top cause accuracy ($\text{ACC}_1$) and CSMF accuracy ($\text{ACC}_\text{CSMF}$) for the 1000 data sets sampled for each simulation having no covariate dependence. The highest performing model is indicated by a bolded mean. Note that FARVA and BF perform quite similarly with regards to both $\text{ACC}_1$ and $\text{ACC}_\text{CSMF}$, as expected. See Table \ref{sim_table} forc data structure for each simulation setting.}
\end{table}

Simulation \textit{c} fixes a common symptom mean structure across causes and symptom covariance structure specific to each cause. In other words, all of the information differentiating CODs is found in the covariance structure. In this setting, models making the conditional independence assumption have no information by which to make a COD determination, and as expected FARVA and BF are the only models to perform better than simple uniform guessing. Simulation \textit{d} fixes a symptom mean and covariance structure specific to each cause. In other words, information about the COD may be found in both the mean and the covariance structure of symptoms. As expected, FARVA and BF outperformed the models that only utilize information about symptom means. Also, FARVA and BF outperformed their own metrics in previous simulation runs, illustrating that both the mean and the covariance have the potential to contain valuable information.

\subsubsection{Simulations with covariate dependence}

Simulation \textit{e} fixes cause-specific symptom mean structure that varies with covariates for each cause and shared independent covariance structure across causes. This simulation is designed to mimic a scenario in which the conditionally independent assumption is valid, and COD-specific symptom prevalence is modified by a single binary covariate. FARVA can utilize this covariate information to more accurately capture cause-specific symptom mean structure for all individuals, and as expected it has the highest performance. Simulation \textit{f} fixes a common symptom mean structure across causes and symptom covariance structure that varies with covariates specific to each cause. Again, only FARVA and BF are expected to do better than chance, and of the two only FARVA can utilize this covariate information to more accurately capture cause-specific symptom association structure for all individuals.  Simulation \textit{g} sets both the mean and covariance of symptoms to depend on cause and vary with a binary covariate. This simulation illustrates that FARVA does best when both the mean and the covariance depend on a covariate, and its performance improves when truly continuous data are included in the model as such. 

\begin{table}
	\footnotesize
	\centering
	\begin{tabular}{ cc||ccccc  }
		\toprule
		\multicolumn{2}{c||}{} &\multicolumn{5}{c}{\textbf{VA model}} \\
		\multicolumn{1}{c}{} & \multicolumn{1}{c||}{Simulation} & 
		\multicolumn{1}{c}{FARVA} & 
		\multicolumn{1}{c}{BF} & 
		\multicolumn{1}{c}{InSilico} & 
		\multicolumn{1}{c}{NBC} & 
		\multicolumn{1}{c}{Tariff} \\
		\midrule
		\multirow{2}{*}{\begin{minipage}{1cm}$\text{ACC}_1$\end{minipage}} & \textit{e} & \textbf{0.442} (0.039) & 0.397 (0.036) & 0.332 (0.079) & 0.389 (0.037) & 0.372 (0.037) \\
		& \textit{f} &  \textbf{0.578} (0.038) & 0.496 (0.038) & 0.255 (0.071) & 0.264 (0.032) & 0.258 (0.038) \\
		& \textit{g}$_1$ & \textbf{0.653} (0.035) & 0.552 (0.038) & 0.333 (0.075) & 0.392 (0.036) & 0.375 (0.038) \\
		& \textit{g}$_2$ & \textbf{0.674} (0.035) & | & |& |&  |\\
		& \textit{g}$_3$ & \textbf{0.749} (0.033)  & | & |& |&  |\\
		\midrule
		\multirow{2}{*}{\begin{minipage}{1cm}$\text{ACC}_\text{CSMF}$\end{minipage}} & \textit{e}   & 0.883 (0.010) & 0.881 (0.011) & 0.647 (0.184) & \textbf{0.884} (0.030) & 0.883 (0.031) \\
		& \textit{f} &  \textbf{0.902} (0.013) & 0.894 (0.014) & 0.647 (0.161) & 0.862 (0.037) & 0.809 (0.086) \\
		& \textit{g}$_1$ & \textbf{0.913} (0.013) & 0.902 (0.014) & 0.649 (0.180) & 0.882 (0.030) & 0.882 (0.031) \\
		& \textit{g}$_2$ & \textbf{0.915} (0.013)  & | & |& |&  |\\
		& \textit{g}$_3$ & \textbf{0.924} (0.013) & | & |& |&  |\\
		\bottomrule
	\end{tabular}
	\caption{\label{sim_table_res_2} Above are the mean (SD) of the top cause accuracy ($\text{ACC}_1$) and CSMF accuracy ($\text{ACC}_\text{CSMF}$)  for the 1000 data sets sampled for each simulation having covariate dependence in the mean, covariance, or both. Recall that different inputs are given to FARVA for the $g$ simulations ($g_1$ uses binary inputs, $g_2$ uses mixed inputs, and $g_3$ uses continuous inputs); the results for the non-FARVA models are shared across each $g$. See Table \ref{sim_table} for other information about the data structure for each simulation setting. The highest performing model is indicated by a bolded mean. The proportion of datasets for which FARVA outperformed BF for simulations $e$ through $g_3$ is 0.91, 1.00, 1.00, 1.00, and 1.00 for $\text{ACC}_1$ and 0.61, 0.82, 0.87, 0.88, and 0.94 for $\text{ACC}_\text{CSMF}$, respectively.}
\end{table}

\subsection{PHMRC predictive performance}

\begin{figure}[!htbp]
	\centering
	\includegraphics[width=1.0\textwidth,keepaspectratio]{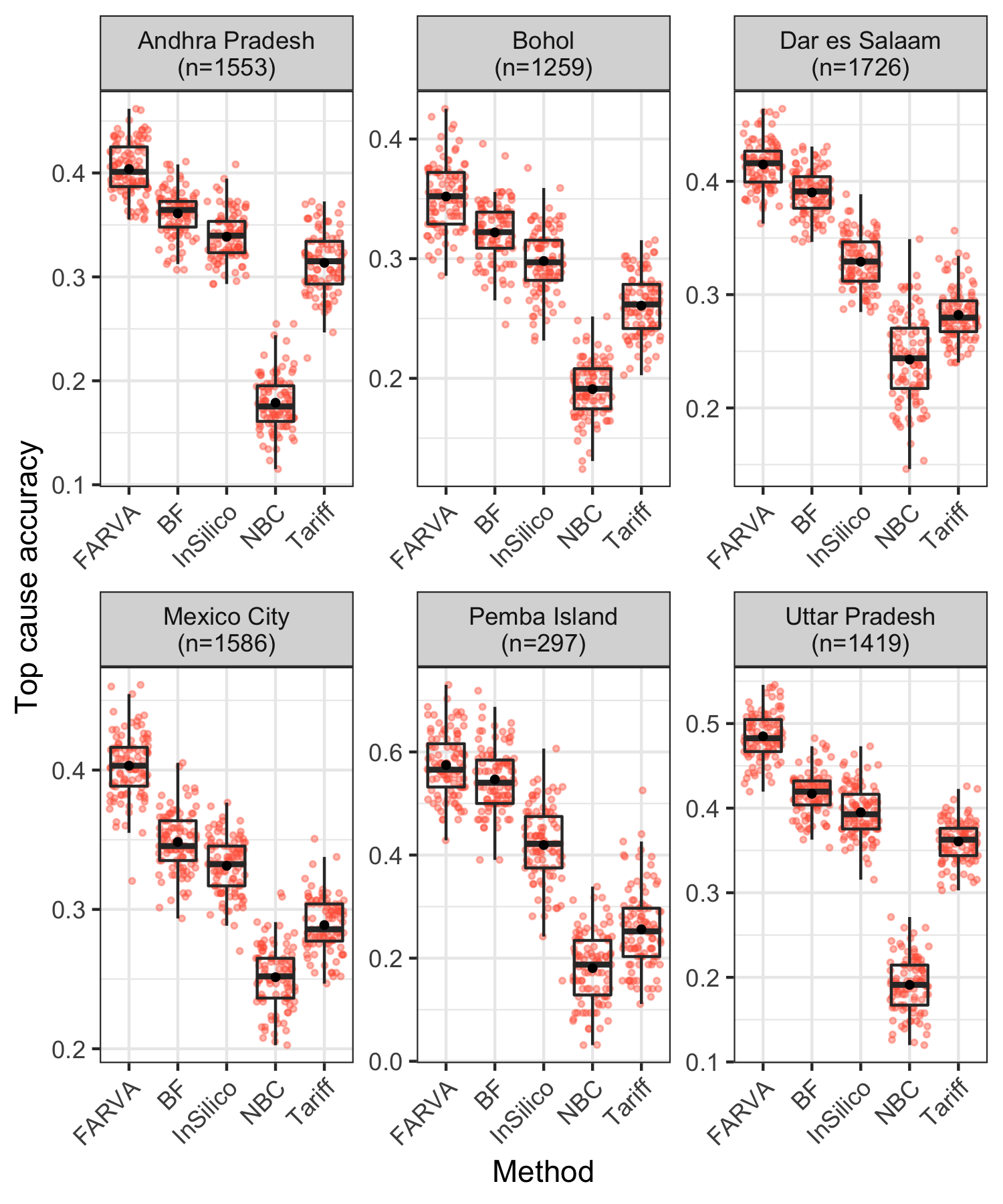}
	\caption{Top cause accuracy for PHMRC data over 100 test/train splits at each site. The centered solid point (horizontal bar) shows the mean (median) across test/train splits, for which individual performance results are shown as semi translucent horizontally jittered points. Each method shares data processing steps, but decedent age ($\geq 65$) is included in FARVA as a binary covariate. For each site (read left to right, top to bottom), the FARVA model outperforms the BF model for 88, 75, 81, 97, 61, and 99\% of the test/train splits. }
	\label{fig:res_phmrc_ACC1}
\end{figure}

Top cause accuracy results for the 100 test/train splits of each site-specific PHMRC data set are shown in Figure \ref{fig:res_phmrc_ACC1}. FARVA outperforms its competitors with regards to top cause accuracy at each site. Notably, FARVA shows improvement relative the BF model in each setting.

\begin{figure}[!htbp]
	\centering
	\includegraphics[width=1.0\textwidth,keepaspectratio]{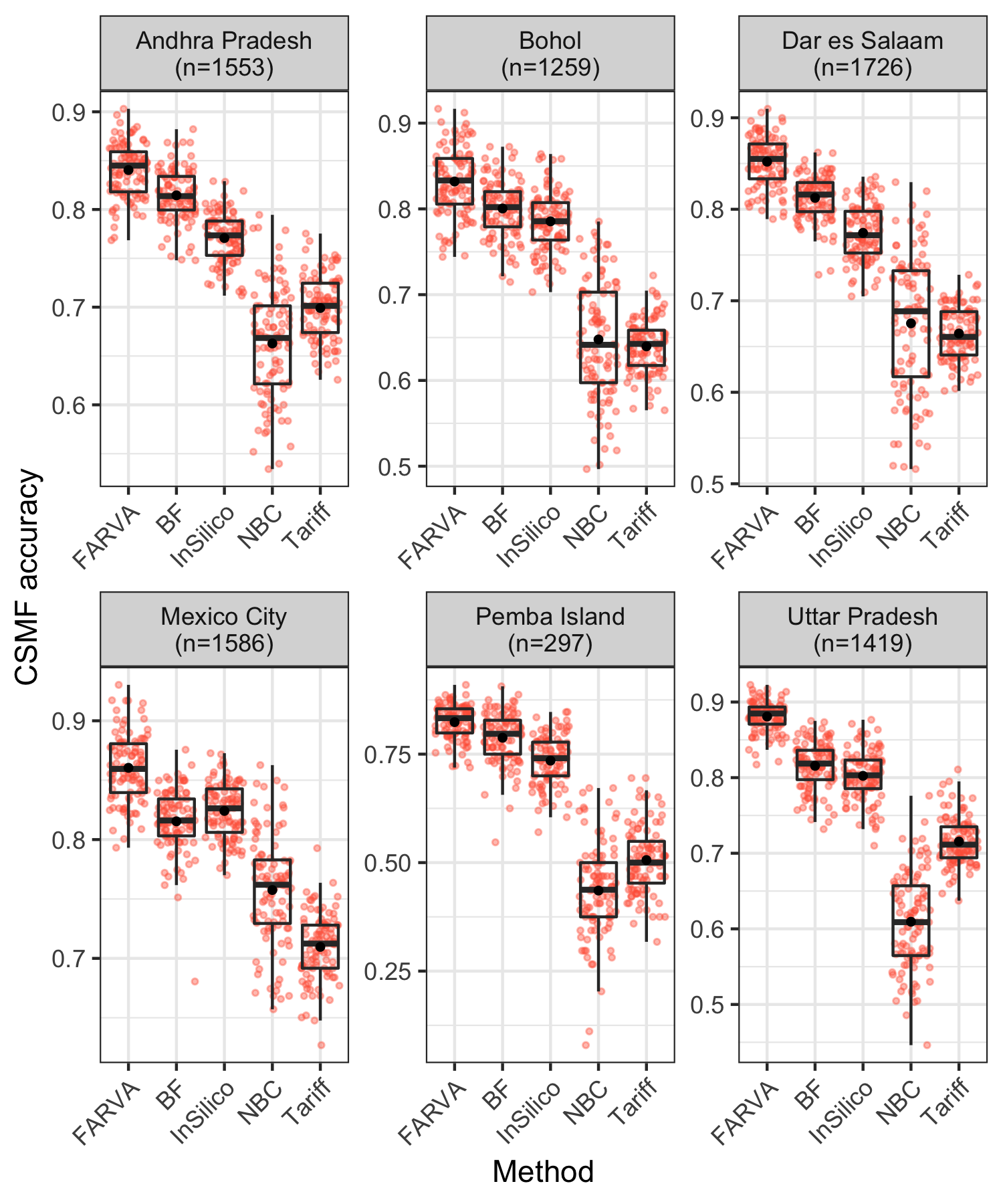}
	\caption{CSMF accuracy for PHMRC data over 100 test/train splits at each site. The centered solid  point (horizontal bar) shows the mean (median) across test/train splits, for which individual performance results are shown as semi translucent horizontally jittered points. Each method shares data processing steps, but decedent age ($\geq 65$) is included in FARVA as a binary covariate. For each site (read left to right, top to bottom), the FARVA model outperforms the BF model for 72, 68, 87, 86, 64, and 97\% of the test/train splits.}
	\label{fig:res_phmrc_CSMF}
\end{figure}

CSMF accuracy results are shown in Figure \ref{fig:res_phmrc_CSMF}. FARVA outperforms its competitors with regards to CSMF accuracy in each site. As before, FARVA shows an improvement in performance over the BF model in each setting. Unsurprisingly, the regions of highest individual performance gain for FARVA are also those with highest CSMF performance gain (e.g., Mexico City and Uttar Pradesh).  However, in some contexts the BF model is able to allocate probability across true causes in test data nearly as well as FARVA (e.g., Pemba Island and to a lesser extent Bohol). See the online supplementary materials for histograms showing the difference in top cause and CSMF accuracy between FARVA and BF in each train/test split


\subsection{Inference on VA questionnaire data}

Of interest in addition to prediction is model-based inference on the structure of responses within VA questionnaires. The FARVA model allows for both the mean and the covariance structure of the latent symptom vector $\bm{z}_i$ to be interrogated. In the following examples the same set of binary symptoms as those used to train the model in the PHMRC predictive ability assessment are considered. The PHMRC variable names and questions are included in full in the online supplementary materials.

\subsubsection{Symptom-level mean}

\begin{figure}[!htb]
	\centering
	\includegraphics[width=0.8\textwidth,keepaspectratio]{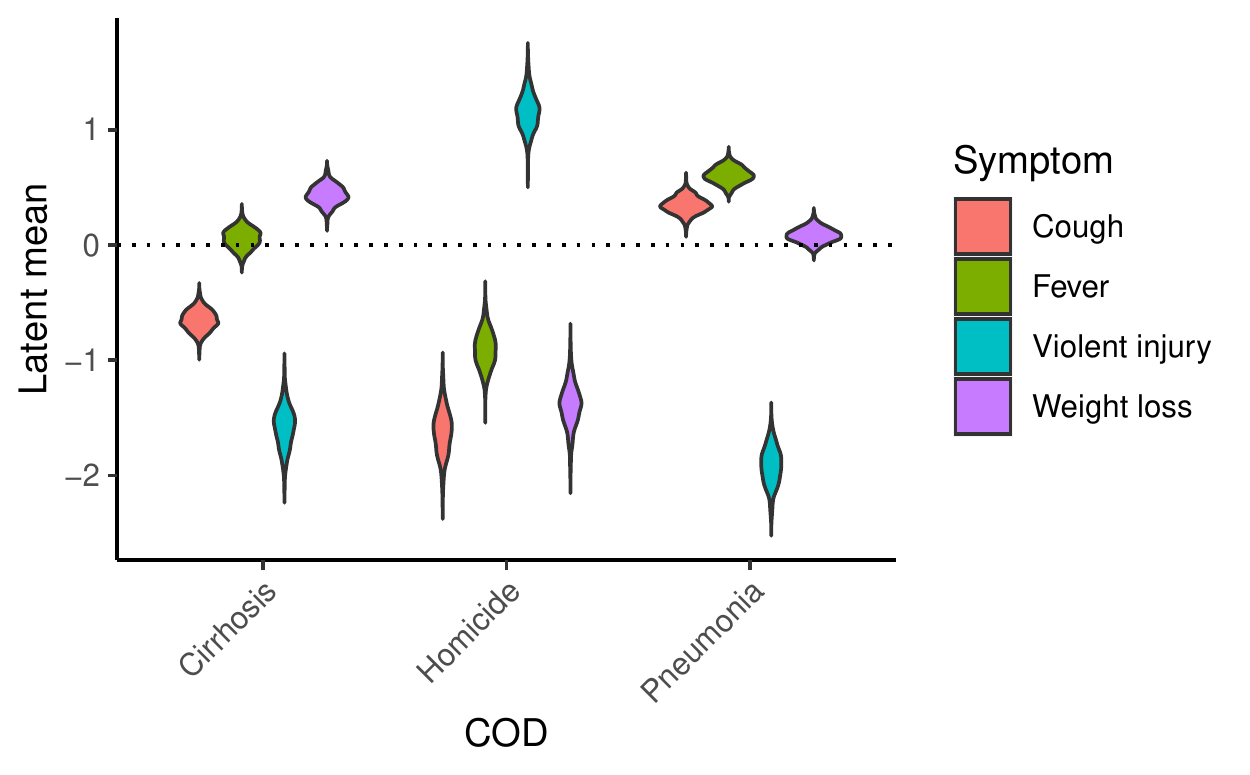}
	\caption{Violin plot visualizing samples of $\mathbb{E}[z_{c[i],j}]$ for CODs cirrhosis ($n=313$), homicide ($n=167$), and pneumonia ($n= 540$) for select symptoms. An expected latent symptom mean being positive (negative) corresponds to more (less) expected observed ``Yes'' responses for those symptoms.} 
\label{fig:inference_full_example}
\end{figure}

First consider an intercept-only FARVA model, i.e. one in which $x_i$ is set to $1$ in equations (\ref{eq:xi_regression}) and (\ref{eq:eta_regression}), run using all 7840 adult deaths in the PHMRC data set. Figure \ref{fig:inference_full_example} illustrates model inference on $\mathbb{E}[\bm{z}_{c[i]}],$ the mean of the latent symptom vector $\bm{z}_{c[i]}$, for some select symptoms and causes. The most important thing to notice about the example is that the latent symptom posterior means differ by cause. For example, samples of $\mathbb{E}[z_{c[i],\text{Cough}}]$ tend to be negative for CODs cirrhosis, and homicide, and positive for COD pneumonia, implying that coughing occurs more frequently for pneumonia deaths. These learned cause-specific patterns of latent symptom means, which differ across CODs, are a part of what drives the model's prediction ability. 

\begin{figure}[!htb]
	\centering
	\includegraphics[width=0.95\textwidth,keepaspectratio]{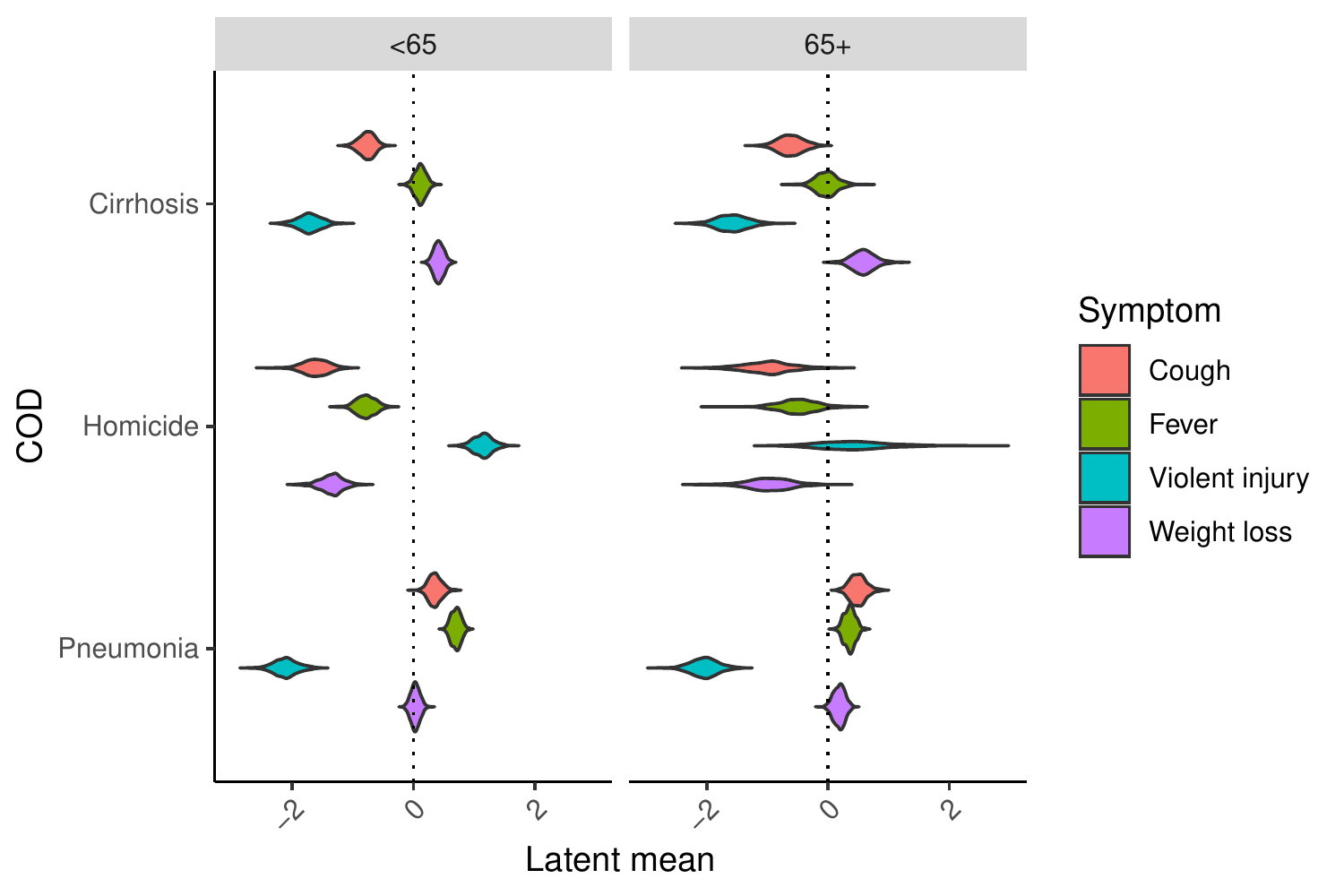}
	\caption{Violin plot of posterior draws of $\mathbb{E}[z_{c[i],j}]$ for CODs cirrhosis ($n_{\geq 65}=49$, $n_{< 65}=260$), homicide ($n_{\geq 65}=5$, $n_{< 65}=161$), and pneumonia ($n_{\geq 65}=193$, $n_{< 65}=341$) for select symptoms.}
	\label{fig:inference_age_example}
\end{figure}

Next consider a FARVA model run using all training data with the binary covariate of  decedent age ($\geq 65$) included, i.e. one in which $\bm{x_i}$ is set to $(\bm{1}_{\text{age}[i]\geq 65},\bm{1}_{\text{age}[i]< 65})'$ in equations (\ref{eq:xi_regression}) and (\ref{eq:eta_regression}). Figure \ref{fig:inference_age_example} provides an example of FARVA's potential to elucidate differences in individual response patterns within COD by covariate, showing $\mathbb{E}[z_{c[i],j}]$ for select symptoms and CODs. The included non-injury-related latent symptom means for homicide deaths are smeared across zero for older decedents, likely due to the small number of observed elder deaths due to homicide combined with the general trend across CODs for senior decedents to experience more constellations of symptoms. Pneumonia appears to present with small differences across age groups for the select symptoms shown, with fever (weight loss) being more (less) common among non-elder decedents. Cirrhosis shows the least age-specific differentiation in these example symptoms.

When age is included as a covariate, it acts as a moderator of the latent mean pattern for a given cause and symptom. The most notable feature of the pattern of symptoms for elders is that they appear less pronounced than symptom patterns of younger decedents (note, e.g., how the 95\% credible interval for cough and fever spans zero for elders even in cirrhosis and homicide CODs). This is consistent with the hypothesis that elders tend to have more diffuse symptoms from a myriad of ailments, not just a pointed set of symptoms leading to a specific diagnosis. The latent mean $\mathbb{E}[z_{c[i],j}]$ is significantly greater than (less than) 0 for 5\% (71\%) of the 4658 possible symptom-cause combinations (137 included symptoms $\times$ 34 causes)  for non-elders, but only 4\% (67\%) of the possible symptom-cause combinations for elders. This phenomenon, although slight, illustrates that FARVA is able to be more focused for younger individuals and spread probability more diffusely for elders, who tend to exhibit a wider array of symptoms across causes, contributing to the improvement in top cause accuracy and CSMF accuracy seen in the previous section. 

\subsubsection{Symptom-level association}

\citet{li2018bayesian} consider inference on overall symptom-level covariance in verbal autopsy data for continuous symptoms, but the results are not cause-specific. \citet{kunihama2018bayesian} discuss a model-based version of Cramer's V, but this metric is only applicable to binary symptoms. To the authors' knowledge, no work has the ability to perform model-based inference on cause-specific covariance in VA data of mixed type.

\begin{figure}[!htb]
	\centering
	\includegraphics[width=0.85\textwidth,keepaspectratio]{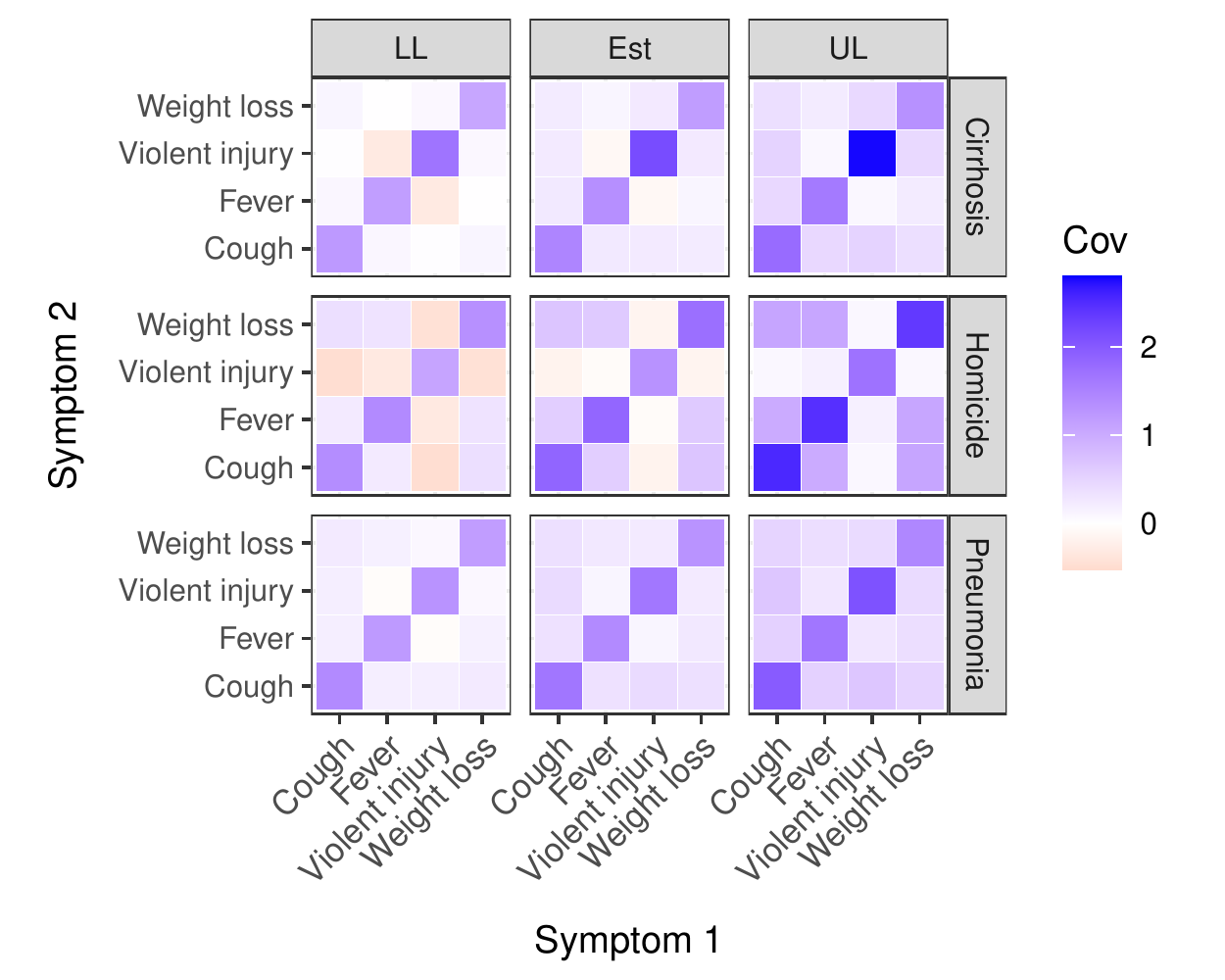}
	\caption{Posterior mean and lower/upper 95\% credible interval of $\text{Cov}(z_{c[i],j},z_{c[i],k})$ for CODs cirrhosis ($n=313$), homicide ($n=167$), and pneumonia ($n= 540$) for select symptoms. An expected latent covariance mean being positive (negative) corresponds to observed ``Yes'' or ``No'' responses for those symptoms to (not) co-occur. For example, samples of $\text{Cov}(z_{c[i],\text{Cough}},z_{c[i],\text{Fever}})$ tend to be positive across all shown CODs, implying that the symptoms cough and fever tend to co-occur. On the other hand, violent injury tends not to be associated with the other symptoms shown, and is negatively associated with these symptoms in the case of homicide deaths.}
	\label{fig:inference_full_example_cov}
\end{figure}

Again, first consider an intercept-only FARVA model run using all training data. Figure \ref{fig:inference_full_example_cov} provides an example of model inference on $\text{Cov}(z_{c[i],j},z_{c[i],k})$ for pairs of symptoms $j$ and $k$, which can be thought of as the latent symptom covariance. These learned patterns of latent symptom covariances capture additional information beyond just symptom prevalence that the model can use to differentiate between CODs.

When age is included as a covariate, it acts as a weak moderator of the latent covariance pattern for given pair of symptoms within a cause (figures included in  online supplementary materials). Of note is that the patterns of association within a COD are similar across age groups, but the strength of association tends to be weaker for younger decedents. Furthermore, the credible intervals tend to be wider for older decedents. This finding differs from the exploratory finding that, across all causes of death, age tends to weaken symptom associations. The difference is due to the conditional nature of these covariance matrices; rather than marginalizing over COD, these symptom associations are conditional on a COD. It is likely that the combination of slightly stronger average covariance along with a wider credible interval is due to the tendency for elderly individuals to be ``hit harder'' by an illness, i.e. to suffer from a large number of the common symptoms for that illness. At the same time, they likely also tend to have many other symptoms unrelated to their COD, and make up a smaller proportion of deaths for each COD, hence the increased uncertainty.

\section{Discussion}\label{disc}

The PHMRC data set, while useful for model testing and validation, has some drawbacks. Deaths clear enough to be considered ``gold standard'' by the PHMRC may not be representative of the population of community deaths, relationships between symptoms/causes learned in one context may not transfer to new contexts (location/time), and, relevant to this work, few potential covariates are recorded. More generally, information in VA data is only as reliable as the interviewee's knowledge. Finally, the question of how to merge different questionnaires (e.g., the adult/child PHMRC form) and select/prioritize symptoms for inquiry remains an open area of research.

For comparability to recent competing methods, the real data analysis of this paper considered only the binary symptoms initially chosen by \citet{byass2012strengthening}. However, the selection of which symptoms to include and whether to include them as binary or continuous variables is an area of open and important research. This issue is not unique to FARVA, and a rigorous consideration of the importance of various symptoms in assigning COD across models would be useful for both model performance and to lower the burden placed on interviewees (a reduced interview set that takes less time would be preferred). An area of potential improvement to the FARVA model itself would be incorporating a prior that could encourage row-level sparsity in the cause-specific loadings matrix $\Lambda_{c[i]}$. This addition could help the model better capture the likely realistic idea that many symptoms may be unimportant to assigning a cause, and could also help assess the relevance of symptoms for inclusion on future questionnaires.

Training on data from one (or multiple) setting(s) and testing in another is common practice for VA algorithms (e.g., using data from Dar es Salaam to train a model, then using test data from Pemba Island). Recent work by \citet{clark2018quantifying} has quantified the effect on predictive performance of this practice. They show that the impact of the information about the joint distribution between symptoms and causes is at least as high as that of the underlying algorithm's logic. In other words, existing algorithms cannot correct for discrepancy between the joint distribution between symptoms and causes in a new site relative to the body of training data. Work by \citet{datta2018local} has focused on accounting for local discrepancy when learning the CSMF at a site with limited training data. It is likely that some (although almost certainly not all) of the variation between sites is systematic and related to measurable features such as seasonal difference, endemicity of various pathogens, urbanicity, etc. If so, some of this decline in predictive performance could be corrected for explicitly in a model like FARVA via the inclusion of region-specific or individual-level covariates in a model. Exploring which covariates help model transferability is a potential future area of research.

The work of McCormick, Clark, Li, and the rest of their team has been pivotal in universalizing algorithm-assigned COD using VA data. In large part, this is due to his team's excellent documentation and code base. The FARVA team strives to maintain similarly high standards of documentation and usability of the code under hopes that it will be friendly to the broader VA community. Algorithm code and a user manual for the FARVA model are available at \url{https://github.com/kelrenmor/farva}.

\section{Acknowledgments}\label{ack}

The authors are grateful to John A. Crump, Manuela Carugati, Matthew P. Rubach, and Michael J. Maze for introducing us to the problem and inspiring us to think about methodological developments, and for helpful discussions with them and the rest of the Investigating Febrile Deaths in Tanzania (INDITe) team. This research was partly supported by the US National Institutes of Health (R01AI121378) for the INDITe project and by the Department of Energy Computational Science Graduate Fellowship (DE-FG02-97ER25308). The funders had no role in study design, data collection and analysis, decision to publish, or preparation of the manuscript.

\appendix
\numberwithin{equation}{section}
\section{Appendix}\label{appx}

Let $U^*$ denote the group of individuals having unknown COD. For person $i^* \in U^*,$ calculate
\begin{equation}\label{unknownCODsamp}
\pi(y_{i^*}=c | \bm{s}_{i^*}) = \frac{\pi(\bm{s}_{i^*} | y_{i^*}=c)\pi(y_{i^*}=c)}{\Sigma_{c^{'}=1}^C \pi(\bm{s}_{i^*} | y_{i^*}=c^{'})\pi(y_{i^*}=c^{'})}
\end{equation}
for each $c=1,\ldots,C$, and sample from the resulting discrete distribution. The online supplemental materials discuss how this value is calulated in practice, along with the computational expense of Monte Carlo integration over $\bm{\eta}_i$ vs. direct sampling.

Then compute the population distribution of causes for individuals in $U^*$ as:
\begin{equation}\label{popdist}
\text{CSMF}_{U^*} = \bigg(\frac{1}{|U^*|} \sum_{i^* \in U^*} 1(y_{i^*}=1), \ldots, \frac{1}{|U^*|} \sum_{i^* \in U^*} 1(y_{i^*}=C) \bigg).
\end{equation}
The above comprises a sample of the CSMF for the set $U^*.$ Note (\ref{popdist}) is different than the posterior distribution of cause-specific probabilities, which includes individuals having both known and unknown CODs and is given by:
\begin{align}
\begin{split}
\{\Pr(y_i=1), &\ldots, \Pr(y_i=C)\} \ | \ \{\Lambda_{c[i]}\} , \{\bm{\eta}_i\}, \Sigma_0 \\
& \sim \text{Dirichlet}\bigg(a_1 + \frac{1}{N} \sum_{i=1}^N 1(y_{i}=1), \ldots, a_C + \frac{1}{N} \sum_{i=1}^N 1(y_{i}=C) \bigg).
\end{split}
\end{align}


\bibliography{bib_FA}


\pagebreak

\end{document}